\documentclass[lettersize,journal]{IEEEtran}
\newcount\DraftStatus  % 0 suppresses notes to selves in text
\usepackage{comment}
\excludecomment{JournalOnly}  
\includecomment{ConferenceOnly}  
\excludecomment{TulipStyle}
\ifnum\DraftStatus=1
	\usepackage[draft,colorinlistoftodos,color=orange!30]{todonotes}
\else
	\usepackage[disable,colorinlistoftodos,color=blue!30]{todonotes}
\fi 
\makeatletter
 \providecommand\@dotsep{5}
 \def\listtodoname{List of Todos}
 \def\listoftodos{\@starttoc{tdo}\listtodoname}
 \makeatother
\usepackage{color}

\usepackage{graphicx}
\graphicspath{{./figures/}{./graphics/}{./graphics/logos/}}
\usepackage{array}
\usepackage{booktabs}
\usepackage{makecell}
\usepackage{tabularx}
\usepackage{algorithm}
\usepackage{algorithmic}
\usepackage{breqn}
\usepackage{subcaption}
\usepackage{multirow}
\usepackage{psfrag}
\usepackage{url}
\usepackage[colorlinks,citecolor=blue]{hyperref}
\usepackage{cleveref}
\usepackage{booktabs}
\usepackage{rotating}
\usepackage{colortbl}
\usepackage{paralist}
\usepackage{epstopdf}
\usepackage{nag}
\usepackage{microtype}
\usepackage{siunitx}
\usepackage{nicefrac}
 \usepackage{grffile}
\usepackage{fontawesome}
\usepackage{xcolor}
\usepackage{multicol}
\usepackage{wrapfig}
\usepackage{todonotes}
\usepackage{tablefootnote}
\usepackage{threeparttable}
\usepackage{bbm} 
\usepackage{cite}
\usepackage{lipsum}
\usepackage[english]{babel}
\usepackage[pangram]{blindtext}
\usepackage{tikz}
\usetikzlibrary{fit,positioning,arrows.meta,shapes,arrows}
\begin{TulipStyle}
\usepackage[numbers]{natbib}
\usepackage{gitinfo2}
\usepackage{fancyhdr}
\usepackage{mathtools}
\usepackage{amsthm}
\theoremstyle{definition}

\theoremstyle{remark}

\numberwithin{equation}{section}
\newtheorem{proposition}{Proposition}
\newtheorem{theorem}{Theorem}

\newtheorem{definition}{Definition}

\hypersetup
{
    pdfauthor={\gitAuthorName},
    pdfsubject={TULIP Lab},
    pdftitle={},
    pdfkeywords={TULIP Lab, Data Science},
}

\end{TulipStyle}

\usepackage{tikz,xcolor}% Make Orcid icon
\definecolor{lime}{HTML}{A6CE39}
\DeclareRobustCommand{\orcidicon}{%
	\begin{tikzpicture}
		\draw[lime, fill=lime] (0,0) 
		circle [radius=0.16] 
		node[white] {{\fontfamily{qag}\selectfont \tiny ID}};    \draw[white, fill=white] (-0.0625,0.095) 
		circle [radius=0.007];    \end{tikzpicture}
	\hspace{-2mm}}
\foreach \x in {A, ..., Z}{%
	\expandafter\xdef\csname orcid\x\endcsname{\noexpand\href{https://orcid.org/\csname orcidauthor\x\endcsname}{\noexpand\orcidicon}}
}

\newtheorem{definition}{Definition}
\newtheorem{theorem}{Theorem}
\newtheorem{proposition}{Proposition}

\IEEEoverridecommandlockouts
\usepackage{amsmath,amssymb,amsfonts}
\usepackage{algorithmic}
\usepackage{graphicx}
\usepackage{textcomp}
\usepackage{tabularx}  
\usepackage{booktabs}
\usepackage{xcolor}
\usepackage[numbers]{natbib}

\def\BibTeX{{\rm B\kern-.05em{\sc i\kern-.025em b}\kern-.08em
    T\kern-.1667em\lower.7ex\hbox{E}\kern-.125emX}}

\begin{document}

%
%=================================================================
% Preamble which will need to be changed for submission
%
\title{Harnessing Uncertainty for Protection: Quantum Differential Privacy with Quantum Noise
}
\title{Towards A Hybrid Quantum Differential Privacy
}

\author{Baobao Song\orcidA{}, Shiva Raj Pokhrel\orcidB{},~\IEEEmembership{Senior Member,~IEEE}, Athanasios V. Vasilakos\orcidC{},~\IEEEmembership{Senior Member,~IEEE}, Tianqing Zhu\orcidD{},~\IEEEmembership{Member,~IEEE}, Gang Li\orcidE{},~\IEEEmembership{Senior Member,~IEEE}
\thanks{Received 14 July 2024; revised 10 December 2024; accepted 11 January 2025.}
\thanks{Baobao Song is with the Faculty of Engineering and IT, University of Technology Sydney, Sydney, Australia~(email: baobao.song@student.uts.edu.au).}%
\thanks{Shiva Raj Pokhrel and Gang Li (corresponding author) are with 
the Centre for Cyber Security Research and Innovation, Deakin University, Geelong, Australia~(email: shiva.pokhrel@deakin.edu.au;gang.li@deakin.edu.au).}
\thanks{Athanasios V. Vasilakos is with the Department of Networks and Communications, College of Computer Science and Information Technology, Imam Abdulrahman Bin Faisal University, Dammam, Saudi Arabia, 
and also with the Department of ICT, Center for AI Research (CAIR), University of Agder, Grimstad, Norway~(th.vasilakos@gmail.com).}
%	 the Department of ICT, Center for AI Research(CAIR), University of Agder, Grimstad, Norway~(email: thanos.vasilakos@uia.no).}%
\thanks{Tianqing Zhu is with the Faculty of Data Science, City University of Macau, Macau, China~(email: Tqzhu@cityu.edu.mo).}}%

%
%\author{\IEEEauthorblockN{Baobao Song}
%\IEEEauthorblockA{\textit{Faculty of Engineering and IT} \\
%\textit{University of Technology Sydney, Melbourne, Australia}\\
%baobao.song@student.uts.edu.au}
%\and
%\IEEEauthorblockN{Shiva Raj Pokhrel}
%\IEEEauthorblockA{\textit{School of Information Technology} \\
%	\textit{Deakin University, Geelong, Australia}\\
%	shiva.pokhrel@deakin.edu.au}
%\and
%\IEEEauthorblockN{Athanasios V. Vasilakos}
%\IEEEauthorblockA{\textit{Department of ICT} \\
%	\textit{University of Agder, Grimstad, Norway}\\
%	thanos.vasilakos@uia.no}
%\and
%\IEEEauthorblockN{Tianqing Zhu}
%\IEEEauthorblockA{\textit{Faculty of Data Science} \\
%	\textit{City University of Macau, Macau, China}\\
%	Tqzhu@cityu.edu.mo}
%\and
%\IEEEauthorblockN{Gang Li}
%\IEEEauthorblockA{\textit{School of Information Technology} \\
%\textit{Deakin University, Geelong, Australia}\\
%gang.li@deakin.edu.au}
%}

\maketitle

\begin{abstract}

% The advent of quantum computing has brought with it 
% the promise of unparalleled processing power, 
% but also introduces challenges in safeguarding data privacy. 
% As the potential for privacy breaches grows, 
% \emph{Quantum Differential Privacy} (QDP) is proposed to ensure the privacy of quantum data.
% Compared to \emph{Differential Privacy} (DP), 
% QDP offers the distinct advantage of leveraging the inherent noise in quantum systems 
% to provide privacy protection. 
% We provide comprehensive noise profiles for quantum systems, 
% analyzing which types of quantum noise are suited as sources for QDP.
% Based on a multi-level noise categorization, 
% we then summarize existing QDP mechanisms, 
% pointing out that most mechanisms simulate 
% only single noise sources, 
% which does not reflect real-world scenarios. 
% Therefore, 
% a resilient QDP mechanism is proposed that 
% combines noise from both channels and measurements.  Through simulation experiments, 
% we provide optimal privacy budget allocations in the resilient QDP mechanism.
% This paper aims to provide a comprehensive overview of the current landscape of QDP, 
% highlighting that QDP has remained largely confined to theoretical explorations 
% and that ongoing efforts are necessary to implement it in reality.

\color{black}
Quantum computing offers unparalleled processing power but raises significant data privacy challenges. \emph{Quantum Differential Privacy} (QDP) leverages inherent quantum noise to safeguard privacy, surpassing traditional DP. This paper develops comprehensive noise profiles, identifies noise types beneficial for QDP, and highlights the need for practical implementations beyond theoretical models. Existing QDP mechanisms, limited to single noise sources, fail to reflect the multi-source noise reality of quantum systems. We propose a resilient hybrid QDP mechanism utilizing channel and measurement noise, optimizing privacy budgets to balance privacy and utility. Additionally, we introduce \textit{Lifted Quantum Differential Privacy}, offering enhanced randomness for improved privacy audits and quantum algorithm evaluation.

\end{abstract}


%\begin{IEEEkeywords}
%component, formatting, style, styling, insert
%\end{IEEEkeywords}

%=================================================================

%=================================================================
\section{Introduction}\label{sec-intro}
Quantum computing significantly advances computational capabilities, 
performing exponentially faster computations than classical computers~\cite{nielsen2001quantum}. 
Quantum algorithms, such as Shor's for factoring large integers~\cite{shor1994algorithms} 
and Grover's for searching unsorted databases~\cite{grover1996fast}, 
highlight its immense power and efficiency. 
This rapid data processing ability~\cite{wittek2014quantum} introduces challenges, 
particularly concerning data privacy~\cite{li2023differential}. 
Differential Privacy (DP) protects individual privacy in databases by adding controlled noise, 
ensuring that any single individual's data does not significantly affect the output~\cite{dwork2006differential}. 
This concept is extended to Quantum Differential Privacy (QDP)~\cite{zhou2017differential}, 
to safeguard quantum privacy by utilizing quantum noise (during transmission and processing). 
Fortunately, the intrinsic noise in quantum systems 
can naturally serve as DP noise~\cite{hirche2023quantum}, 
potentially eliminating the need for artificial noise.
{\color{black} Our research focuses on developing robust QDP models to ensure privacy in the quantum era, addressing challenges like noise from decoherence and imperfect operations. We aim to improve the practical applicability of QDP by designing models effective in realistic quantum systems, overcoming limitations such as uncontrolled environmental noise~\cite{du2021quantum}. Additionally, we explore harnessing intrinsic quantum noise to reduce dependency on artificial noise, enhancing efficiency~\cite{hirche2023quantum}. Our work addresses gaps in literature by reviewing and clarifying relationships among existing QDP mechanisms and proposes resilient, flexible models adaptable to diverse scenarios. Finally, we focus on reducing sample complexity by redefining privacy testing frameworks for quantum systems, enabling efficient and reliable data privacy protection.}

{\color{black} We harness quantum noise to meet QDP requirements and simulate realistic quantum systems accurately. Our work develops comprehensive noise profiles by identifying types, sources, and magnitudes of noise affecting quantum operations and channels. We classify noise types suitable for QDP and summarize existing mechanisms, proposing a resilient QDP mechanism to simulate the cumulative effects of multi-source noise. Finally, we introduce \textit{Lifted Quantum Differential Privacy (Lifted QDP)}, extending QDP by incorporating randomized quantum datasets and rejection sets, addressing key limitations of traditional QDP.

The main contributions of this paper are as follows: we develop \textit{comprehensive noise profiles for quantum systems, identifying noise types that can be leveraged for QDP} and \textit{summarizing existing QDP mechanisms}. A simple and \textit{resilient QDP mechanism} is proposed to simulate multi-source noise in realistic quantum systems, optimizing privacy budget allocation. Additionally, we \textit{design the proof-of-concept of Lifted QDP}, which enhances flexibility, reduces sample complexity, and enables the reuse of quantum datasets and models for multiple tests. The proposed hybrid QDP mechanism ensures effective noise management and balances privacy and utility in quantum data processing, while Lifted QDP introduces improved flexibility and adaptability.

\color{black}
\subsection{Preliminaries and Concepts}\label{sec-preliminaries}
{\textcolor{black} {In this section, we introduce the key concepts of DP and QDP.
%		
% definition and basic mechanism of DP. We then present QDP, a state-of-the-art approach for protecting individual data privacy in quantum systems. 
Then, we present Lifted DP, which enables more effective statistical tests for privacy audits.
DP~\cite{dwork2006differential} protects individual privacy by adding noise to data, ensuring personal information remains undisclosed. Neighboring datasets differ by only one individual's data.}}
DP is defined as:

%\subsubsection{Privacy Models}\label{sec-dpm}

\begin{definition}\label{def:dp}
    \textbf{$(\epsilon,\delta)$-Differential Privacy}~\cite{dwork2006differential}. 
    A random algorithm $\mathcal{M}$ provides $(\epsilon,\delta)$-differential privacy if for any two datasets $D$ and $D'$ differing by at most one record, and for all outputs $A \in Range(O)$:
    \begin{equation}\label{equ:4}
        Pr[\mathcal{M}(D) \in O] \leq e^\epsilon Pr[\mathcal{M}(D') \in O] + \delta;
    \end{equation}
    where, $\epsilon$ controls the privacy level, and $\delta$ accounts for the probability that $\epsilon$ does not hold.
\end{definition}

Local Differential Privacy (LDP) ensures privacy by adding noise directly at the data source:
\begin{definition}\label{def:ldp}
    \textbf{$(\epsilon,\delta)$-LDP}~\cite{dwork2014algorithmic}. 
    A local algorithm $\mathcal{M}$ provides $(\epsilon,\delta)$-local DP if for any two values $x$ and $x'$, and for every output $y$:
    \begin{equation}\label{equ:5}
        Pr[\mathcal{M}(x) = y] \leq e^\epsilon Pr[\mathcal{M}(x') = y] + \delta.
    \end{equation}
\end{definition}

\subsubsection{Differential Privacy Mechanisms}\label{sec-dpme}

DP mechanisms adjust the probability distribution of query results 
to mask individual contributions. 
Key mechanisms include the \textit{Laplace Mechanism} 
and \textit{Gaussian Mechanism}~\cite{liu2018generalized}:

\begin{definition}\label{def:lm}
    \textbf{Laplace Mechanism}. 
    Given a function $f: D \rightarrow \mathbb{R}^d$, the Laplace mechanism is defined as:
    \begin{equation}\label{equ:6}
        M_{Lap}(D) = f(D) + (X_1, X_2, \ldots, X_d).
    \end{equation}
    $X_i$ are i.i.d random variables from the Laplace distribution with $\mu = 0$ and $\sigma = \frac{\Delta_g f}{\epsilon}$, where $\Delta_g f = \max \limits_{\lVert D - D' \rVert_1} \lVert f(D) - f(D') \rVert_1$.
\end{definition}

\begin{definition}\label{def:gm}
    \textbf{Gaussian Mechanism}. 
    Given a function $f: D \rightarrow \mathbb{R}^d$, the Gaussian mechanism is defined as:
    \begin{equation}\label{equ:7}
        M_{Gau}(D) = f(D) + (X_1, X_2, \ldots, X_d).
    \end{equation}
    $X_i$ are i.i.d random variables from the Gaussian distribution with $\mu = 0$ and $\sigma = \frac{\Delta_g f \sqrt{2 \log(\frac{1.25}{\delta})}}{\epsilon}$.
\end{definition}

\textit{Randomized Response} is used in LDP to collect sensitive data while preserving privacy:
\begin{definition}\label{def:rr}
    \textbf{Randomized Response}. 
    Given a query $f$ for a binary sensitive attribute $A$, the randomized response mechanism ensures:
    \begin{equation}\label{equ:8}
        A' = 
        \begin{cases} 
            A & \text{with probability } p = \frac{e^\epsilon}{e^\epsilon + 1}, \\
            1 - A & \text{with probability } 1 - p = \frac{1}{e^\epsilon + 1}.
        \end{cases}
    \end{equation}
\end{definition}

\subsubsection{DP Properties}\label{sec-dpp}

DP and LDP satisfy three fundamental properties~\cite{dwork2014algorithmic},
which are shown below:

\begin{proposition}[Post-processing]\label{pro1}
	If a mechanism $M$ provides $\epsilon,\delta$-DP, then any function $f$ applied to the output of $M$ provides at most $\epsilon,\delta$-DP. 
\end{proposition}
\color{black}{
\begin{proposition}[Parallel composition]\label{pro2}
If independent algorithm $M_1$ and $M_2$ satisfy $(\epsilon_1, \delta_1)$-DP and $(\epsilon_2, \delta_2)$-DP, respectively, and , then $M_1 \otimes M_2$ is $(\epsilon_1 + \epsilon_2, \delta_1 + \delta_2)$-DP.
\end{proposition}

%
%\begin{proposition}[Parallelism]\label{pro3}
%	If we have two datasets $D_1, D_2,..., D_k$, 
%	and we apply DP mechanisms $M_1$, $M_2$,..., and $M_k$ to each dataset respectively, 
%	then  for $i \in [k]$, 
%	the combined mechanism $M_i$ applied to $D_i$ in parallel 
%	will provide the same level of privacy as applying $M_1, M_2,... M_k$ sequentially.
%\end{proposition}

%\subsection{Quantum Differential Privacy} \label{sec-ddp}

\begin{table}[h]
	\centering\scriptsize
	\caption{The detailed comparison between DP and QDP }
	\label{comparisonqdp}
	\begin{tabular}{>{\centering\arraybackslash}m{1.5cm} >{\centering\arraybackslash}m{2.5cm} >{\centering\arraybackslash}m{3.3cm}} 
		\toprule % Top horizontal line
		\textbf{Aspect} & \textbf{Differential Privacy (DP)} & \textbf{Quantum Differential Privacy (QDP)} \\ % Column names
		\midrule [0.2pt]% Middle horizontal line
		Protection Objectives & Individual record & Personal data during quantum computation and transformation processes \\ 
        \midrule[0.2pt]
		Neighbors in Definition & Two datasets that they differ by one record & Two quantum states are close in quantum metrics like trace distance \\ 
  		\midrule[0.2pt]
		Key Features & Mathematical techniques to add randomness to data or queries & Quantum mechanics principles like superposition to enhance  privacy \\
  		\midrule[0.2pt]
		Three properties& \checkmark & \checkmark \\
		\midrule[0.2pt]
		Applications& Public statistical data releases, research involving sensitive data & Quantum data processing, secure quantum communications, and quantum cryptography \\
		\bottomrule % Bottom horizontal line
	\end{tabular}
\end{table}
% %\emph{Quantum Qifferential privacy} (QDP)  aims to achieve differential privacy protection through quantum computing technology.
% Specifically, 
% QDP introduces quantum noise during quantum computations to protect individual privacy.
% At its core, 
% QDP inherits the foundational concepts of classical DP,
% such as privacy-preserving data analysis 
% and the quantification of privacy loss. 
% It adapts these concepts to the unique properties 
% and operations of quantum information, 
% leveraging quantum mechanics to enhance privacy protection in quantum settings.
\color{black}
A comparison of DP and QDP in various aspects are shown in \Cref{comparisonqdp}. In QDP, 
neighboring quantum states are defined
based on the concept of distance in the quantum state space~\cite{hirche2023quantum}. 
Two quantum states are considered neighbors
if they are close to each other in terms of a chosen distance metric, 
such as the trace distance,
which is the sum of the absolute values of the differences
    between the eigenvalues of the two states. }
\color{black}
\begin{definition} \label{def:td}	\textbf{Trace Distance}. 
	For two quantum states $\rho$ and $\sigma$, 
	the trace distance is given by
	\begin{equation}\label{equ:8}
		||\rho - \sigma||_1 = \frac{1}{2} \sum_i |\lambda_i(\rho) - \lambda_i(\sigma)|.
	\end{equation}
	where $\lambda_i(\rho)$ and $\lambda_i(\sigma)$ are the eigenvalues of $\rho$ and $\sigma$, respectively.
\end{definition}

Intuitively, 
the trace distance measures how distinguishable two quantum states are. 
When the trace distance between two states is small, 
it indicates that the states are similar or close to each other. 
Conversely, 
a large trace distance indicates that the states are significantly different.
In the context of QDP, 
the trace distance is used to define the notion of neighboring quantum states, 
and neighboring quantum states can be expressed as:

\begin{definition} \label{def:nqs}	\textbf{Neighboring Quantum States}. 
	Given two quantum states $\rho$ and $\sigma$,
	they are considered neighboring 
	if their trace distance is below a certain threshold $d$:
	\begin{equation}\label{equ:9}
		||\rho - \sigma||_1 \leq d .
	\end{equation}
\end{definition}

Then, QDP can be defined as:
\begin{definition} \label{def:qdp}	\textbf{Quantum Differential Privacy}. 
	A quantum mechanism satisfies $(\epsilon, \delta)$-QDP 
	if for any pair of neighboring quantum states $\rho$ and $\sigma$, 
	and for any quantum measurement $M$, 
	the following condition holds:
	\begin{equation}\label{equ:10}
		Tr[M(\rho) \in S] \leq e^{\epsilon} \cdot Tr[M(\sigma) \in S] + \delta
	\end{equation}
	where $Tr[M(\rho) \in S]$ and $Tr[M(\sigma) \in S]$ are the probabilities 
	of obtaining measurement outcomes in a certain set $S$ 
	when measuring the states $\rho$ and $\sigma$ respectively. 
	%$\epsilon$ quantifies the privacy loss, 
	%and $\delta$ represents the additional privacy loss beyond the target $\epsilon$.
\end{definition}

This definition ensures that the probability ratio of 
obtaining measurement outcomes for neighboring quantum states 
is bounded by $e^\epsilon$ with an additional small term $\delta$, 
providing a probabilistic guarantee of privacy protection.
\textcolor{black}{
QDP still satisfies the post-processing property (\Cref{pro1}),
but for parallel composition, 
$M_1 \otimes M_2$ satisfies $ (\epsilon, \delta) $-QDP, 
where $ \epsilon = \epsilon_1 + \epsilon_2$
and $\delta = \min \left\{ \delta_1 + e^{\epsilon_1} \delta_2, \delta_2 + e^{\epsilon_2} \delta_1 \right\}$~\cite{hirche2023quantum, angrisani2023unifying}.}

\color{black}
%\subsection{Lifted Differential Privacy} \label{sec-lifteddp}

From an auditing perspective, 
DP mechanisms need carefully calibrated noise to protect privacy, 
but verifying that a given implementation truly meets privacy guarantees is challenging. 
Auditing DP often involves generating counterexamples (like canaries) to test for privacy breaches. 
Traditional DP can be inflexible, 
as it relies on fixed datasets and single statistical tests, 
limiting the audit process. 
Auditors have to train models from scratch for each trial, 
and since each test is tied to a specific dataset, 
the sample complexity (the number of model runs needed to spot privacy issues) can be very high.

Therefore, Lifted Differential Privacy (Lifted DP)~\cite{pillutla2024unleashing} was proposed. 
Lifted DP expands upon DP by allowing randomized datasets and rejection sets, 
which offers greater flexibility in the auditing process. 
Randomized dataset variations in Lifted DP enable more generalized counterexamples 
to be constructed using randomly sampled canaries, 
which can be tested against.
This not only leads to more accurate privacy guarantees but also makes the auditing process more practical and scalable.

\begin{definition} \label{def:lidp}	\textbf{Lifted Differential Privacy}. 
A randomized mechanism $ \mathcal{A} $ satisfies $(\epsilon, \delta)$-Lifted Differential Privacy if, 
for random neighboring dataset pairs $D_0$ and $D_1 $
and any randomized rejection set $R$, the following holds::
	\begin{equation}\label{equ:30}
	Pr[\mathcal{A}(D_1) \in R] \leq e^{\epsilon} Pr[\mathcal{A}(D_0) \in R] + \delta
	\end{equation}
where $Pr$ considers both the randomness of the mechanism and the dataset distribution. 
\end{definition}

\color{black}

\subsection{Quantum Noise for QDP}\label{sec-noise}

Noise is unavoidable in quantum systems, as shown in \Cref{fig:noisesore}, 
affecting qubit states and computation results 
due to various disturbances and errors~\cite{soare2014experimental}. 
This noise leads to errors and information loss. 
However, noise enables natural privacy barrier in quantum computing~\cite{ju2024quantum}. 
Recent studies suggest modeling noise to meet QDP criteria~\cite{huang2023certified,bai2024quantum}. 
% We examine the nature of noise in quantum systems, 
% provides comprehensive noise profiles, 
% and explores its potential for QDP. 
% Quantum noise are of four types: 
% \textit{coherent noise, decoherence, crosstalk, and measurement noise}. 
% The relationship between these noise types 
% and their physical origins is shown in \Cref{fig:noisesore} 
% and discussed in the following.

\begin{figure}[htbp]
	\centering
	\includegraphics[width = 0.669\linewidth]{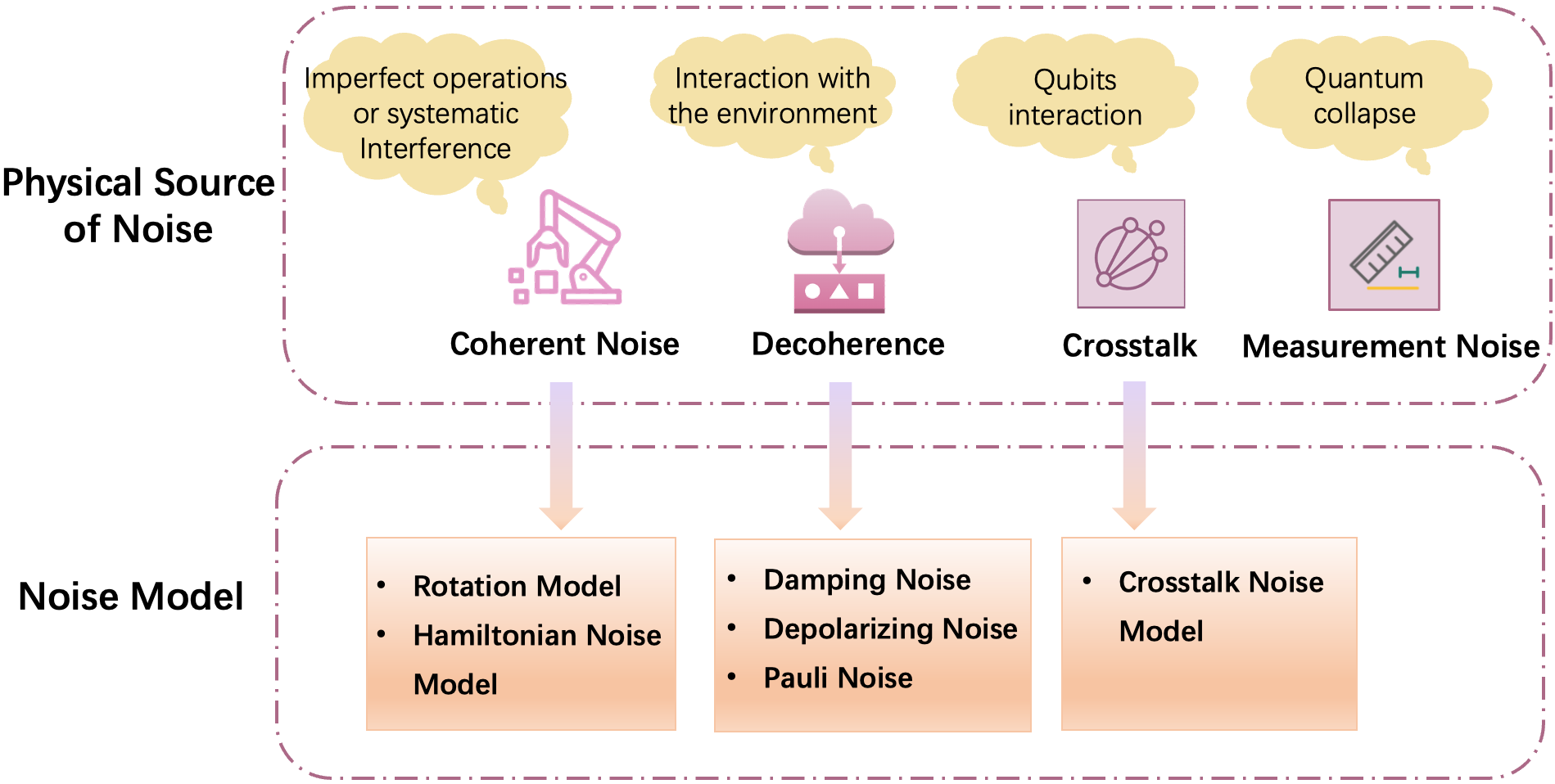}
	\caption{Quantum noise sources and noise models}
	\label{fig:noisesore}
\end{figure}

\emph{Coherent Noise}\label{sec-coherent}. Quantum coherence describes the ability of a quantum system 
to maintain a phase relationship among its states~\cite{li2016quantum}, 
where phase is the angular component 
of the complex probability amplitude that characterizes these states.
Quantum coherence allows quantum systems to exist in 
superposition and entangled states, 
which are essential for quantum computing to effectively process information.
Coherent noise refers to a type of noise that, 
while maintaining the coherence of quantum states, 
introduces systematic errors~\cite{penshin2024realization}. 
It mainly originates from inaccuracies in the control of quantum systems, 
device imperfections, 
and specific environmental fluctuations~\cite{ahsan2022quantum}.
An example of coherent noise is presented in \Cref{fig:coherent}, 
where it can be observed that the noise induces a rotation of the quantum state on the Bloch sphere, 
with the green rotation axis perpendicular to the plane of rotation.

\begin{figure}[!htbp]
	\centering
	\begin{subfigure}[b]{0.17202\textwidth}
		\centering
		\includegraphics[width=\textwidth]{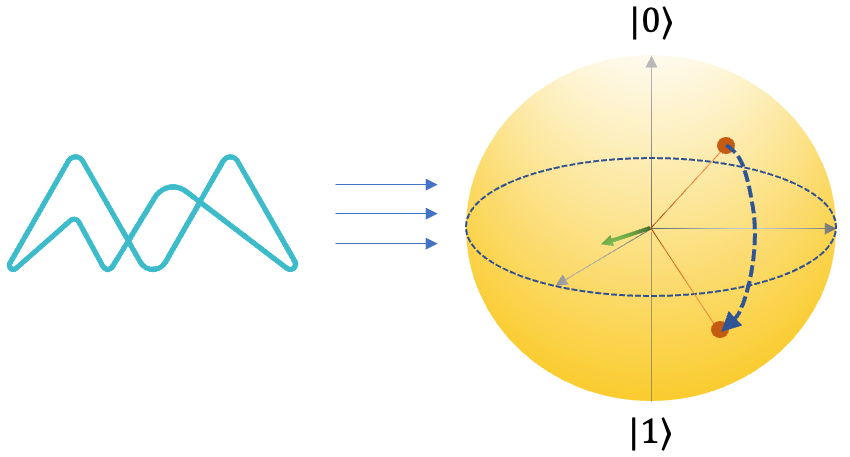}
		\caption{Coherent noise}
		\label{fig:coherent}
	\end{subfigure}
	\hspace{0.7cm}  %add space
	\begin{subfigure}[b]{0.15182\textwidth}
		\centering
		\includegraphics[width=\textwidth]{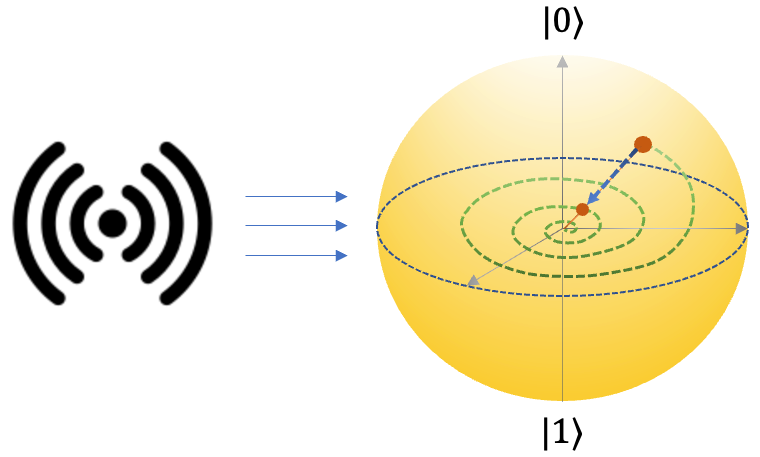}
		\caption{Decoherence}
		\label{fig:deco}
	\end{subfigure}
	\caption{
		A quantum state affected by coherent noise and decoherence, with arrows showing state changes.
%	In (a), 
%	the qubit remains on Bloch sphere surface, 
%	while in (b), 
%	the qubit ends up inside the Bloch sphere.
}
	\label{fig:coandede}
\end{figure}

Since coherent noise typically manifests as systematic and deterministic errors, 
such as consistent phase shifts introduced in each operation, 
it can be to some extent anticipated and corrected~\cite{greenbaum2017modeling}. 
Therefore, 
naturally occurring coherent noise is not suitable as a noise source for QDP, 
as it can potentially be predicted or reversed by attackers. 
Nevertheless, 
a noise model that simulates coherent noise can be employed to 
introduce artificial noise into quantum states, 
thereby making it challenging for attackers to predict the noise. 
Rotation models, 
such as Z-rotations~\cite{hu2021css} 
and X-rotations with Pauli-X errors~\cite{greenbaum2017modeling}, 
have been designed to simulate coherent noise 
and they have been demonstrated to achieve QDP. 
The existing proposed mechanisms related to this approach are discussed in \Cref{enqs}.
Another model for coherent noise is the Hamiltonian noise model~\cite{zaborniak2021benchmarking}, 
which reflects the effect of noise on quantum state evolution 
by changing the Hamiltonian of the system. 
%The introduction of a stochastic Hamiltonian in quantum systems is an effective way to add noise 
%that can be used in QDP
%but currently, Hamiltonian noise model is not used in QDP. 
%Future research may explore QDP mechanisms tailored to this model, 
%possibly involving the randomization of Hamiltonian matrix elements
%or the definition of stochastic Hamiltonians through rotational noise.

\emph{Decoherence}\label{sec-decoherence}. Decoherence refers to the process by which a quantum system loses its coherence, 
usually due to perturbations from the external environment~\cite{paz2002environment}, 
such as thermal noise. 
This process leads to the disruption of the superposition 
and entanglement properties of qubits
resulting in the irreversible loss of quantum information.
As shown in \Cref{fig:deco}, 
decoherence causes a qubits move from the surface of the sphere (representing a pure state) 
toward the center of the sphere (representing the maximally mixed state). 
The blue arrow indicates the trajectory of the qubit's movement.

The noise that causes decoherence is random as the interactions 
between the quantum system and its environment 
are complex~\cite{korcyl2011studies}. 
This type of noise is well-suited to QDP since its randomness 
can effectively obscure the information 
in the original quantum state. 
Phase damping model~\cite{pirandola2008minimal}, 
amplitude damping model~\cite{srikanth2008squeezed}, 
depolarizing model~\cite{collins2015depolarizing}, 
and Pauli model~\cite{chen2022quantum} 
can effectively simulate the evolution of quantum systems 
following their interactions with the environment. 
Their corresponding QDP mechanisms have also been proposed, 
as detailed in \Cref{inqc}. 
However, decoherence noise is 
inherently challenging to regulate with precision. 
Models of decoherence are mainly theoretical, 
and thus do not accurately reflect the characteristics of 
decoherence noise in the real world. 
Consequently, the effectiveness of QDP mechanisms based on decoherence in protecting privacy 
remains theoretical and requires further investigation to validate.

\emph{Crosstalk}\label{sec-crosstalk}. In quantum systems,  
operations on one qubit can unintentionally affect 
the state of neighboring qubits,
a phenomenon known as crosstalk~\cite{zhou2023quantum}.
Crosstalk arises due to the physical proximity and coupling between qubits. 
% As illustrated in \Cref{fig:crosstalk}, 
% crosstalk frequently results in the generation of 
% uncontrollable noise effects on adjacent qubits.

% \begin{figure}[htbp]
% 	\centering
% 	\includegraphics[width = 0.346\linewidth]{figure/crosstalk2.png}
% 	\caption{Crosstalk noise on qubits}
% 	\label{fig:crosstalk}
% \end{figure}

Crosstalks are random but not independent.
Its origin lies in the interactions between qubits, 
and thus it does not meet the independence requirement for noise in QDP. 
Nevertheless, 
in special cases, DP also have correlated noise. 
For example, 
in time series, 
correlated data points necessitate the presence of associated noise 
in order to accommodate the inherent properties of the data~\cite{wang2021current}.
Another example is that carefully designed correlated noise 
can not only provide privacy protection 
but also reduce the impact of accumulated noise on gradients~\cite{koloskova2023gradient}.
Therefore, 
although there has not yet been any research 
on utilizing crosstalk to implement QDP, 
by carefully designing appropriate crosstalk noise or 
by combining other models to increase the randomness of crosstalk, 
it is possible that it could also become one of the noise sources for QDP.

\emph{Measurement Noise}\label{sec-measurement}. Quantum measurement is not merely a passive reading of information,
but an active disturbance of the quantum system~\cite{ozawa2004uncertainty}.
The interaction between the measurement device and the qubits causes the qubits' states to change,
known as quantum state collapse.
After quantum measurement, 
the quantum state indeed collapses to either 
$|1 \rangle$  or  $|0 \rangle$,
determined by the probability amplitudes of the superposition state before measurement. 
\Cref{fig:measure}  illustrates the collapse possibility of a quantum state in measurement.
Quantum measurement noise sources 
such as insufficient sensitivity of measurement devices, detector noise, 
and uncertainty in quantum state collapse~\cite{drossel2018contextual}. 

\begin{figure}[htbp]
	\centering
	\includegraphics[width = 0.46\linewidth]{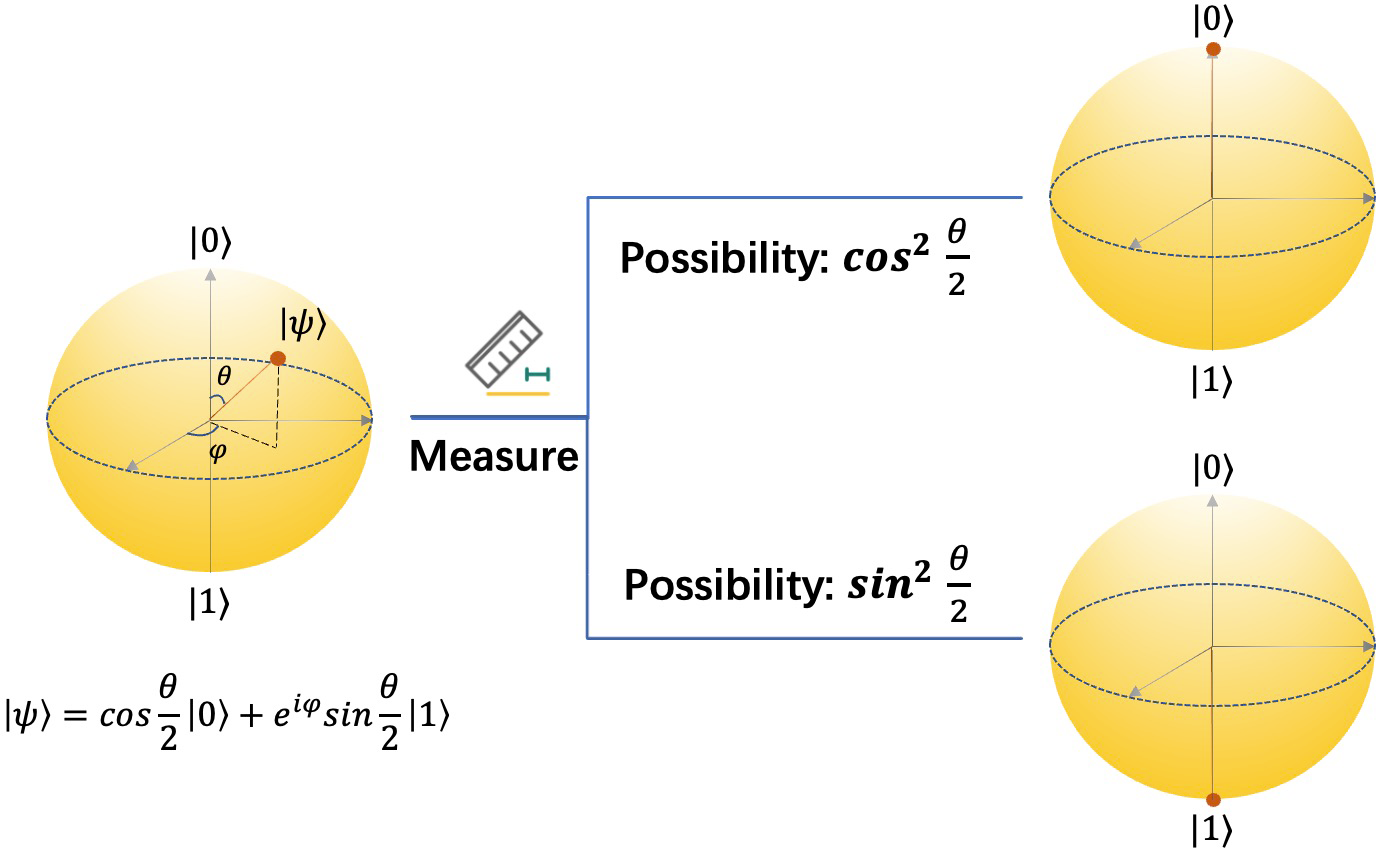}
	\caption{The measure result of a qubit $| \psi \rangle$}
	\label{fig:measure}
\end{figure}

The collapse process is inherently random 
and unpredictable,
which renders the associated measurement noise
likewise random, unpredictable, 
and independent of the quantum state, 
meeting the noise criteria for QDP. 
However, 
quantum measurement noise is closely related to 
the specific measurement devices and processes used 
(e.g., the noise from identical equipment is consistent)
and in order to improve measurement accuracy, 
multiple measurements are often performed to
obtain a more precise estimate of the quantum state~\cite{wecker2015progress}. 
Multiple measurements allow for the estimation of noise introduced by the equipment 
and enable a more accurate determination of the quantum state 
based on the probabilities of state collapse, 
effectively mitigating the impact of noise.
Therefore, 
limited times of measurement are adopted to realize QDP,
which is shown in \Cref{inm}. 
\Cref{comparisonnoise} lists the quantum noise discussed in this section 
and compares their suitability as noise sources for QDP theoretically.

\begin{table}[h]
	\renewcommand{\arraystretch}{0.9}
	\centering\scriptsize
	\caption{The comparison of quantum noise for realizing QDP}
	\label{comparisonnoise}
	\begin{tabular}{>{\centering\arraybackslash}m{1.2cm} >{\centering\arraybackslash}m{1.7cm} >{\centering\arraybackslash}m{1.3cm}>{\centering\arraybackslash}m{1cm}>{\centering\arraybackslash}m{1.4cm}} 
		\toprule [1pt]% Top horizontal line
		\textbf{Aspect} & \textbf{Coherent noise} & \textbf{Decoherence}& \textbf{Crosstalk}& \textbf{Measurement noise} \\ % Column names
		\midrule % Middle horizontal line
		Independent & \checkmark & \checkmark&$\times$& \checkmark  \\ \midrule
		Randomness & $\times$ & \checkmark & \checkmark & \checkmark  \\ \midrule
		Controllable & \checkmark &  \checkmark &  $\times$ &\checkmark \\ \midrule
		Suitable for QDP & Adopt the coherent noise model to add noise artificially & Use the channel noise& Hard to  design related model& Limit the times of measurement  \\ 
		\bottomrule % Bottom horizontal line
	\end{tabular}
\end{table}

%=================================================================
\section{Quantum Differential Privacy Mechanisms}\label{sec-methods}

%To prevent individual privacy from being inferred from query statistics, 
%DP is used to protect personal information. 
%This is primarily achieved by employing various DP mechanisms 
%that introduce random noise into the output results 
%to obscure the true data. 
%This process can also be extended to the quantum domain, 
%where the noise is introduced during quantum computation to protect private data. 
With the analysis of quantum noise in \Cref{sec-noise}, 
this section categorizes the existing QDP mechanisms based on a hierarchical categorization scheme in \Cref{fig:tree}.  
As the non-independent nature of crosstalk noise, 
which makes it difficult to utilize, 
existing QDP mechanisms mainly focus on  following noise: 
coherent noise, decoherence, and measurement noise.

As shown in \Cref{fig:tree}, 
the first layer (blue blocks) categorizes noise based on its attributes, 
specifically whether the noise is an inherent internal noise
or an artificially introduced external noise.
%identifying inherent internal noise and externally added noise 
%as the two primary sources for achieving QDP. 
%
%
%In QDP, 
%the noise introduced to preserve privacy originates from two main sources: 
%inherent noise and external noise. 
Inherent noise is a natural phenomenon in quantum systems 
due to several fundamental and unavoidable factors, 
primarily including interactions with the environment, 
the intrinsic properties of quantum mechanics, 
and imperfections in the quantum hardware~\cite{benenti2019principals}.
%These various sources of inherent noise pose significant challenges 
%for maintaining the stability and reliability of quantum computation. 
%At the same time, 
%it has been shown that this inherent noise is intrinsically compatible with the mechanisms of QDP, 
%providing natural privacy protection for data~\cite{hirche2023quantum}.
In contrast, external noise,
distinct from naturally occurring disturbances, 
is deliberately introduced to enhance privacy protection. 
%This noise is strategically injected into the quantum system to 
%obscure the specific contributions of individual data points, 
%ensuring that the output of a quantum algorithm does not reveal sensitive information 
%about any single entry in the dataset. 
In the second layer (green blocks), 
we further differentiate the mechanisms 
based on the specific location where the noise is introduced. 
Finally, in the third layer (yellow blocks),
we present the specific mechanisms implemented.
From the analysis presented in \Cref{sec-noise}, 
it can be concluded that decoherence 
and the measurement noise resulting from quantum collapse 
are naturally occurring phenomena, 
so they belong to internal noise.
In addition,  
the coherent noise simulated through noise models is considered external noise. 
The introduction of artificial noise during the measurement phase also categorizes as external noise.
\Cref{sec-inherent} and \Cref{sec-external} discuss privacy protection mechanisms 
based on internal and external noises, respectively, 
\Cref{hybrid} proposes a resilient QDP mechanism 
which combines the different types of noise discussed above
and make simulations to provide the optimal privacy budget allocation.
\Cref{summarym} provides a summary for QDP mechanisms.

%
%
%One thing to note is that we regard quantum channels as sources of internal noise. 
%Quantum channels describe the state changes of qubits 
%during operations or transmission within quantum circuits, 
%including potential noise and interference~\cite{blais2020quantum}. 
%Although many quantum channels are artificially constructed, 
%they are used to study and analyze the behavior and performance of quantum systems in the presence of noise. 
%While it is possible to artificially introduce  quantum channels to add noise,  
%the essence of these channels is to simulate the errors 
%that qubits may encounter during transmission or processing in reality. 
%Therefore, in this paper, 
%various quantum channels such as depolarizing channel and pauli channel 
%are considered of inherent noise in quantum computing.

\begin{figure}[htbp]
	\centering
	\includegraphics[width = 0.7\linewidth]{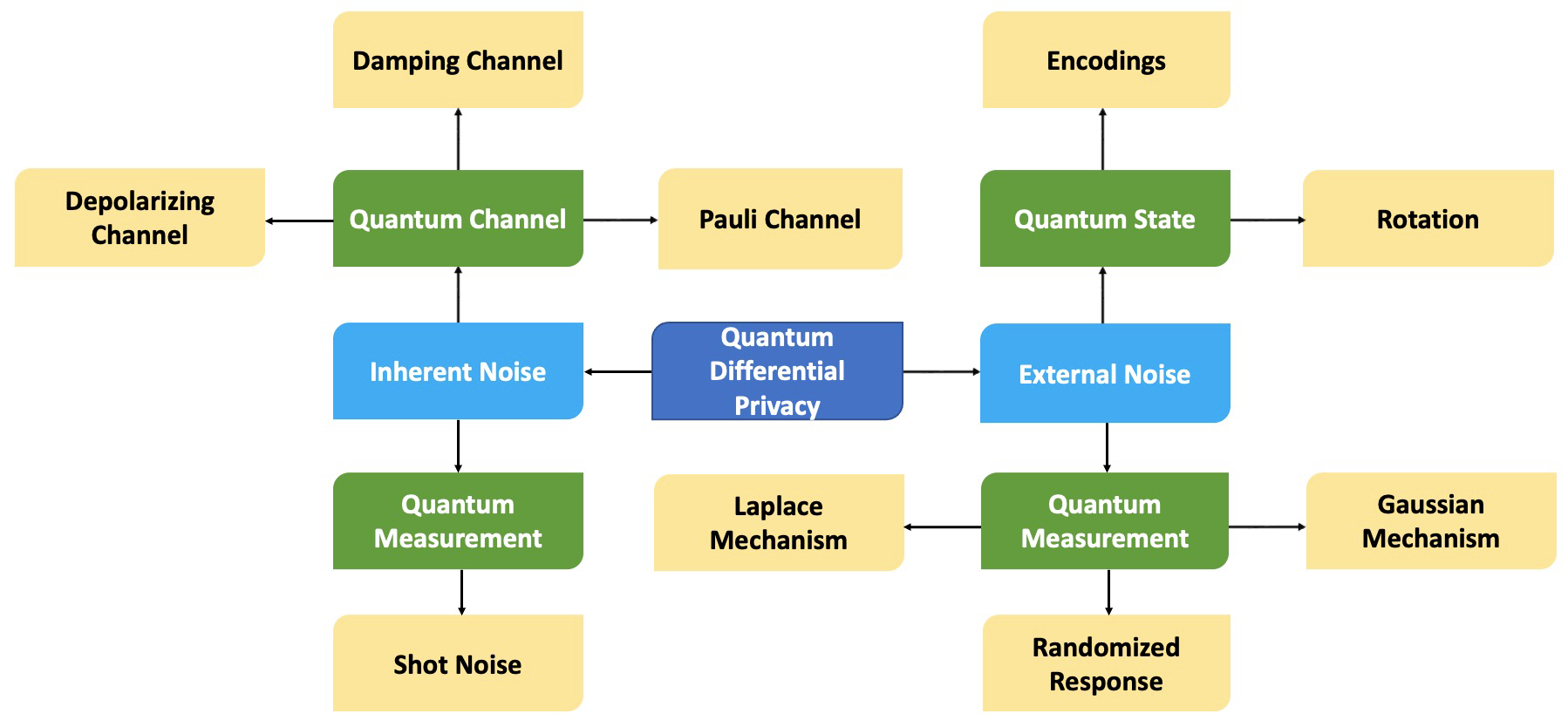}
	\caption{High-level view of QDP}
	\label{fig:tree}
\end{figure}

\subsection{Inherent Noise to Realize QDP}\label{sec-inherent}
In quantum computing, 
inherent noise refers to the unavoidable disturbances 
that affect quantum systems during computation. 
These noises are intrinsic to the physical systems~\cite{shimizu2002stability}. 
The primary source of inherent noise is decoherence~\cite{schlosshauer2019quantum},  
which commonly occurs in quantum channels 
and can be modeled as depolarizing, damping, 
and Pauli channels. 
These are introduced in \Cref{inqc}. 
\Cref{inm} discusses the internal noise 
generated by a limited number of measurements in quantum systems, 
known as shot noise~\cite{li2023differential}.

\subsubsection{Inherent Noise on Quantum Channel}\label{inqc}
Inherent noise in quantum channels, 
such as depolarizing, damping, and Pauli channels, 
significantly impacts quantum systems. 
The depolarizing channel introduces noise by randomly flipping qubit states, 
causing them to depolarize to mixed states, 
which is fundamental for quantum error correction and differential privacy. 
Damping channels, including amplitude and phase damping, 
simulate energy loss from qubits to their environment, 
modeling real-world scenarios where quantum information gradually degrades. 
Pauli channels apply random Pauli operators (X, Y, Z) to qubits, 
introducing various types of error. 
As discussed in the following, 
understanding these noise channels is crucial for developing robust QDP mechanisms 
to protect and preserve the integrity of quantum data.

\emph{Depolarizing Channel}. The depolarizing noise channel is the most extensively studied aspect for QDP~\cite{zhou2017differential, du2021quantum}.
It describes that qubits randomly lose their original quantum state information 
during transmission or processing due to random noise
and tend towards a completely mixed state.
%
%the random errors that qubits have during transmission or processing 
%due to external environmental factors. 
Specially, 
depolarizing channel assumes that a qubit has a certain probability of undergoing a completely random state change: 
\begin{definition} \label{def:dc}	\textbf{Depolarizing  Channel}. 
	\begin{equation}\label{equ:11}
		E_{\text{Dep}}(\rho) = (1 - p)\rho + \frac{p}{2}I
	\end{equation}
	where $\rho$ is the original state of the qubit, 
	$p$ is the depolarizing probability, 
	and $I$ is the identity matrix. 
	This formula describes how the channel transforms the qubit state 
	into a mixture of the original state 
	and the maximally mixed state, 
	ensuring that with probability $p$, 
	the state is completely randomized.
\end{definition}

From \Cref{equ:11}, 
it can be seen that the depolarizing noise channel 
masks the influence of individual quantum states 
by randomly transforming the quantum state into a mixed state, 
thereby protecting privacy. 
This random noise is not only unpredictable 
but also independent of the quantum state, 
making it a suitable source of noise for quantum QDP.
% Therefore, 
% \citet{zhou2017differential} first proposed depolarizing mechanism that satisfies QDP
% and they presented the theoretical derivation,
% which demonstrates the differential privacy parameter formula 
% for the depolarizing mechanism. 

\begin{theorem}[Depolarizing Mechanism~\cite{zhou2017differential}]. \label{the:1} 
	Given the  trace distance $d$ and the dimension of the Hilbert space $D$, 
	depolarizing mechanism provides $\epsilon$-QDP where
	\begin{equation}\label{equ:12}
		\epsilon_{dep} = \ln \left[ 1 + \frac{(1 - p)dD}{p} \right]
	\end{equation}
\end{theorem}

%Given the  trace distance $d$ and the dimension of the Hilbert space $D$, 
%the $\epsilon$ of depolarizing mechanism is: 
%
%\begin{equation}\label{equ:12}
%\epsilon = \ln \left[ 1 + \frac{(1 - p)dD}{p} \right]
%\end{equation}

\Cref{the:1} provided the theoretical foundation 
for using depolarizing channels to ensure privacy in quantum computing, 
which has been applied to provide protection for quantum algorithms.
For example, 
\citet{yang2023improved} presented 
privacy-preserving \emph{Hybrid Quantum-Classical Algorithms}, 
which employed the depolarizing mechanism to 
make the quantum circuits meet the QDP requirement.
\citet{du2021quantum} proposed a algorithm 
to enhance the robustness of quantum classifiers 
against adversarial attacks by leveraging depolarization noise 
into the quantum circuits. 
They provided theoretical guarantees for the robustness of 
quantum classifiers with added depolarization noise in the context of sampling.
%studied both inherent noise and external noise, 
%using combinatorial theory to demonstrate the advantages of natural quantum noise. 
%They also employed a depolarizing mechanism 
%to provide differential privacy protection.

There are also efforts focused on 
providing tighter bounds for the $\epsilon$  parameter 
to ensure that depolarizing channels meet the QDP.
\citet{bai2024quantum} calculated $\epsilon$ under the depolarizing channel
base on the proportional distance,
a metric used to calculate the extent of change 
between two quantum states after 
they pass through a quantum channel. 
They gave a better differential privacy parameter $\epsilon=\ln(1 + d)$
of the depolarizing mechanism than \Cref{equ:12} in some constrains. 
Moreover, 
some works introduce the concept of \emph{hockey-stick divergence}
for achieving QDP~\cite{hirche2023quantum, angrisani2023unifying}
and provide new bounds. 
\citet{hirche2023quantum} redefined QDP using quantum \emph{hockey-stick divergence}, 
that is under the influence of the quantum algorithm,
the quantum \emph{hockey-stick divergence}
$E_{e^\epsilon}$ between adjacent states 
does not exceed a predetermined privacy bound $\delta$. 
They utilized the properties of quantum \emph{hockey-stick divergence} 
such as data processing inequality and contraction coefficients
to derive the new bound of global and locally depolarizing mechanism,
as illustrated in~\Cref{comparisonchannel}. 
Following \citet{hirche2023quantum} work,
\citet{angrisani2023unifying} proved the \emph{advanced joint convexity of the quantum hockey-stick divergence}, 
which describes the divergence behavior of quantum states after passing through a quantum channel, 
and used it to update the privacy bounds of generalized noisy channel.

%
%\begin{definition} \label{def:ajc}	\textbf{Advanced joint convexity of the quantum hockey-stick divergence}
%	\begin{equation}\label{equ:18}
	%E_{\gamma_1}(p \rho_0 + (1 - p) \rho_1 \| p \rho_0 + (1 - p) \rho_2) \leq (1 - p)(1 - \beta) E_\gamma(\rho_1 \| \rho_0) + (1 - p) \beta E_\gamma(\rho_1 \| \rho_2)
	%	\end{equation}
%where $\gamma_1 = 1 + (1 - p)(\gamma - 1)$  and $\beta = \frac{\gamma_1}{\gamma}$. 
%$\rho_0, \rho_1, \rho_2$ are  quantum states.
%\end{definition}
%
%In essence, \cref{equ:18} shows how the divergence 
%between convex combinations of quantum states 
%(i.e., weighted sums) can be bounded by 
%the divergences between the individual states.
%It helps in analyzing the privacy guarantees of quantum channels and mechanisms.
%By understanding how the divergences between quantum states combine, 
%they derive tighter privacy bounds for global and local depolarizing channel as below:
%
%\begin{equation}\label{equ:19}
%\delta \leq \max \left\{ 0, (1 - \exp(\epsilon)) \frac{p}{2^n} + (1 - p) d \right\}
%\end{equation}
%
%\begin{equation}\label{equ:20}
%\delta \leq \max \left\{ 0, (1 - \exp(\epsilon)) \frac{p^k}{2^k} + (1 - p^k) d \right\}
%\end{equation}

~\\
\emph{2. Damping Channel}

The damping channel is a common type of quantum noise channel 
in quantum computing
used to describe the interaction of a quantum system with its environment, 
resulting in the loss of energy~\cite{d2012transmission}. 
%It captures the decoherence effects that occur 
%due to spontaneous emission and other dissipative processes.
\emph{Generalized Amplitude Damping Channel} (GAD) ~\cite{srikanth2008squeezed} is a one of the most common damping channels, 
%The main types of damping channels are the \textit{Generalized Amplitude Damping Channel} (GAD) ~\cite{srikanth2008squeezed}
%and the \textit{Phase-Amplitude Damping Channel} (PAD)~\cite{li2023quantum}. 
which is typically used to simulate energy exchange process
in a quantum system interacting with a thermal equilibrium environment
and  describes the changes in quantum states. 
As \emph{Kraus operators} provide a way to represent quantum channels,
describing the evolution of the state of a quantum system 
under the influence of an environment or a noisy process~\cite{rajagopal2010kraus}, 
the definition of GAD with \emph{Kraus operators} can be represented by:

\begin{definition} \label{def:gad}	\textbf{Generalized Amplitude Damping Channel}
	\begin{equation}\label{equ:13}
		\mathcal{E}_{\text{GAD}}(\rho) = \sum_{k=0}^{3} E_k \rho E_k^\dagger
	\end{equation}
	with the probability $\gamma$  representing the energy exchange between the system and the environment 
	and the probability $p$ that the environment is in the ground state, 
	the Kraus operators $E_k$ are defined as:
	\begin{equation}\label{equ:14}
		E_0 = \sqrt{p} \begin{pmatrix}
			1 & 0 \\
			0 & \sqrt{1 - \gamma}
		\end{pmatrix}, \quad 
		E_1 = \sqrt{p} \begin{pmatrix}
			0 & \sqrt{\gamma} \\
			0 & 0
		\end{pmatrix}
	\end{equation}
	\begin{equation}\label{equ:15}
		E_2 = \sqrt{1 - p} \begin{pmatrix}
			\sqrt{1 - \gamma} & 0 \\
			0 & 1
		\end{pmatrix}, \quad 
		E_3 = \sqrt{1 - p} \begin{pmatrix}
			0 & 0 \\
			\sqrt{\gamma} & 0
		\end{pmatrix}
	\end{equation}
\end{definition}

In GAD, 
the parameters $p$ and $\gamma$ allow for fine-tuning of the random noise levels,
which is crucial for QDP to provide privacy protection. 
By adding controlled level of GAD noise, 
it becomes difficult for attackers to distinguish between similar quantum states, 
thereby safeguarding sensitive information and maintaining data utility.
Consequently, 
\citet{zhou2017differential} put forward the  GAD mechanism for QDP. 

\begin{theorem}Generalized Amplitude Damping Mechanism~\cite{zhou2017differential}.\label{the:2} 
	For all inputs $\rho$ and $\sigma$ 
	satisfying $\tau(\rho, \sigma) \leq d$, 
	the GAD noise mechanism $\mathcal{M}_{\text{GAD}}$ 
	provides $\epsilon$-differential privacy, where
	\begin{equation}\label{equ:16}
		\epsilon_{gad} = \ln \left[ 1 + \frac{2d\sqrt{1-\gamma}}{1-\sqrt{1-\gamma}} \right]
	\end{equation}
\end{theorem}

Besides, 
\citet{zhou2017differential} further studied the \emph{Phase-Amplitude Damping Channel} (PAD) , 
which not only considers the energy exchange process between the quantum system and the environment 
but also involves coherence loss. 
Additionally, 
they quantified the relationship between the parameters controlling noise in PAD and $\epsilon$ in QDP.
Based on ~\cite{zhou2017differential},
\citet{angrisani2022differential} 
researched how the combination of quantum channels amplifies QDP, 
and by leveraging the contraction coefficients of these channels, 
he derived the explicit bounds of the combination of depolarizing channel and the GAD 
for the privacy parameter.

%The \textit{Phase-Amplitude Damping Channel} (PAD) 
%is often used to simulate the behavior of 
%photons scattering randomly through a waveguide. 
%It captures both the loss of energy (amplitude damping) 
%and the loss of coherence (phase damping) in a quantum system.
%The operation of the PAD can be represented 
%by the following Kraus operators:
%
%\begin{definition} \label{def:pad}	\textbf{Phase-Amplitude Damping Channel}
%	\begin{equation}\label{equ:25}
	%\mathcal{E}_{\text{PAD}}(\rho) = \sum_{k=0}^{3} F_k \rho F_k^\dagger
	%	\end{equation}
%	where the Kraus operators $F_k$ are defined as:
%	\begin{equation}\label{equ:26}
	%		F_0 = \begin{pmatrix}
		%			1 & 0 \\
		%			0 & \sqrt{1 - \lambda}
		%		\end{pmatrix}, \quad 
	%		F_1 = \begin{pmatrix}
		%			0 & \sqrt{\lambda} \\
		%			0 & 0
		%		\end{pmatrix}
	%	\end{equation}
%\end{definition}
%
%The PAD noise mechanism for QDP also proposed by \citet{zhou2017differential}.
%For parameters $\gamma$ and $\lambda$ satisfying $\lambda \leq \gamma$, 
%the PAD noise mechanism $\mathcal{M}_{\text{PAD}}$ provides $\epsilon$-differential privacy, where
%
%\begin{equation}\label{equ:25}
%\epsilon = \ln \left[ 1 + \frac{2d \sqrt{1 - \gamma} \sqrt{1 - \lambda}}{1 - \sqrt{1 - \gamma} \sqrt{1 - \lambda}} \right].
%\end{equation}

~\\
\emph{3. Pauli Channel}

The Pauli channel is used to model various types of noise 
that can occur during the transmission or storage 
of quantum information~\cite{chen2022quantum}. 
It is named after the Pauli matrices, 
which represent the simple yet comprehensive models of phase-flip
and bit-flip errors that can affect qubits.
The action of the Pauli channel on a quantum state 
$\rho$ can be described as:

\begin{definition} \label{def:pauli}	\textbf{Pauli Channel}
	\begin{equation}\label{equ:17}
		\mathcal{E}_{\text{Pauli}}(\rho) = p_I I \rho I + p_x \sigma_x \rho \sigma_x + p_y \sigma_y \rho \sigma_y + p_z \sigma_z \rho \sigma_z
	\end{equation}
	where 
	\begin{equation}\label{equ:18}
		\sigma_x = \begin{pmatrix}
			0 & 1 \\
			1 & 0
		\end{pmatrix}, \quad
		\sigma_y = \begin{pmatrix}
			0 & -i \\
			i & 0
		\end{pmatrix}, \quad
	\end{equation}
	\begin{equation}\label{equ:19}
		\sigma_z = \begin{pmatrix}
			1 & 0 \\
			0 & -1
		\end{pmatrix},  \quad
		I = \begin{pmatrix}
			1 & 0 \\
			0 & 1
		\end{pmatrix}, 
	\end{equation}
	where $p_I$, $p_x$, $p_y$, and $p_z$ are the probabilities 
	of applying the identity operation, 
	Pauli-X matrix (bit-flip) $\sigma_x$, 
	Pauli-Y matrix (combination of bit-flip and phase-flip operations) $\sigma_y$, and Pauli-Z matrix (phase-flip) $\sigma_z$ respectively.
	These probabilities satisfy the condition $p_I + p_x + p_y + p_z = 1$.
\end{definition}

Because the different Pauli matrices in the Pauli channel 
represent spin-flip operations of qubits in different directions, 
the Pauli channel can be used to simulate qubits' behavior 
during transmission under various noise sources. 
By appropriately selecting the probabilities of the Pauli operators, 
it is possible to introduce a controllable level of random noise in the Pauli channel, 
thereby achieving the desired level of privacy protection. 
\citet{bai2024quantum} studied two special cases of the Pauli channel: 
the phase-flip channel and the bit-flip channel.
When $p_x=0$ and $p_y=0$, 
the Pauli channel is equivalent to the phase-flip channel.
Similarly,
when $p_y=0$ and $p_z=0$, the Pauli channel simplifies to the bit-flip channel. 
\citet{bai2024quantum} proposed specific methods and privacy bounds 
for achieving QDP in quantum computing 
using the phase-flip channel and the bit-flip channel. 
Additionally, 
by analyzing the decay rate of the trace distance 
(change in the trace distance between quantum states after passing through a quantum channel) 
and fidelity (protective performance of different quantum channels on quantum states), 
they demonstrated that, 
under the same privacy budget, 
the depolarizing channel exhibits a higher privacy protection capability 
compared to the phase-flip channel and the bit-flip channel. 
\Cref{comparisonchannel} lists the comparison of all the aforementioned channel mechanisms 
that satisfy DQP along with their privacy bounds.

%\citet{angrisani2023unifying} discussed Pauli noise as part of a broad noise channel 
%to ensure differential privacy in quantum algorithms.
%They also employed the \emph{advanced joint convexity of the quantum hockey-stick divergence}
%to derive the privacy guarantee bounds of QDP for the Pauli channel.
%
%\begin{theorem}\label{the:1} 
%	Given the  trace distance $d$ and the dimension of the Hilbert space $D$, 
%	depolarizing mechanism provides $\epsilon$-QDP where
%	\begin{equation}\label{equ:12}
	%		\epsilon = \ln \left[ 1 + \frac{(1 - p)dD}{p} \right]
	%	\end{equation}
%\end{theorem}

\begin{table}[htbp]
	\caption{The comparison of channel models for QDP}
	\centering\scriptsize
	\label{comparisonchannel}
	%\centering
	\resizebox{\columnwidth}{!}{
		\begin{tabular}{c c c}
			\toprule
			\textbf{Quantum Channel} & \textbf{Source of Noise} & \textbf{Privacy Parameters} \\ 
			\midrule
			\multirow{4}{2.5cm}{\centering Depolarizing  Channel} & 
			\multirow{4}{5cm}{\centering External environmental interference and imperfections within the system} 
			& $\epsilon_{dep} = \ln \left[ 1 + \frac{(1 - p)dD}{p} \right]$~\cite{zhou2017differential} \\ & &$\epsilon_{dep}=\ln(1 + d)$ for $d \in (0,1]$~\cite{bai2024quantum} \\ 
			&&	$ \epsilon_{global}= \max \left\{ 0, \ln \left( 1 + \frac{D}{p} \left( (1 - p)d - \delta \right) \right) \right\}\textsuperscript{1}$ ~\cite{hirche2023quantum}\\ 
			&&$\epsilon_{local} = \max \left\{ 0, \ln \left( 1 + \frac{D^k}{p^k} \left( (1 - p^k)d - \delta \right) \right) \right\}\textsuperscript{1}$~\cite{hirche2023quantum}\\
			\midrule
			\multirow{2}{2.5cm}{\centering Damping Channel } & 
			\multirow{2}{5cm}{\centering Energy loss, thermal dissipation, spontaneous emission, and others} 
			& $\epsilon_{gad} = \ln \left[ 1 + \frac{2d\sqrt{1-\gamma}}{1-\sqrt{1-\gamma}} \right]$~\cite{zhou2017differential} \\ 
			& &$\epsilon_{pad} = \ln \left[ 1 + \frac{2d \sqrt{1 - \gamma} \sqrt{1 - \lambda}}{1 - \sqrt{1 - \gamma} \sqrt{1 - \lambda}} \right]$~\cite{zhou2017differential} \\ 
			\midrule
			Phase-flip channel & The loss of phase information & 
			$\epsilon_{pf} = 
			\begin{cases} 
				\ln\left(1 + \frac{d}{2(1-p)}\right), & \text{if } 0 \leq p < \frac{1}{2} \\
				\ln\left(1 + \frac{d}{2p}\right), & \text{if } \frac{1}{2} \leq p \leq 1
			\end{cases}$~\cite{bai2024quantum}\\
			\midrule
			Bit-flip channel & The loss of bit information& $\epsilon_{bf}=\ln(1 + d)$~\cite{bai2024quantum}\\
			\midrule
			\multirow{2}{2.5cm}{\centering Generalized noisy channel} & 
			\multirow{2}{5cm}{\centering All types of noise} 
			&$\delta_{global} \leq \max \left\{ 0, (1 - \exp(\epsilon)) \frac{p}{2^n} + (1 - p) d \right\}\textsuperscript{2}$~\cite{angrisani2023unifying}\\
			&&$\delta_{local}  \leq \max \left\{ 0, (1 - \exp(\epsilon)) \frac{p^k}{2^k} + (1 - p^k) d \right\}\textsuperscript{2}$~\cite{angrisani2023unifying} \\ 
			\midrule
			\multirow{1}{2.5cm}{\centering The combination of depolarizing  channel and PAD} & 
			\multirow{1}{5cm}{\centering Random environmental interference, device imperfections, energy loss , thermal excitation, phase decoherence, and others} & 
			\parbox[c]{6cm}{\centering \vspace{4mm} $\epsilon_{dap}=(1 - p) \ln \left( 1 + \frac{2d \sqrt{1 - \gamma} \sqrt{1 - \lambda}}{1 - \sqrt{1 - \gamma} \sqrt{1 - \lambda}} \right)$~\cite{angrisani2022differential}}
			\vspace{9mm}
			\\
			\bottomrule
		\end{tabular}
	}
	\parbox{\columnwidth}{\tiny
		1. $\epsilon_{global}$  and $\epsilon_{local}$ are for global depolarizing mechanism and  local depolarizing mechanism respectively.
		$k$ indicates number of qubits.\\
		2. $\delta_{global}$ and $\delta_{local}$ are for global generalized channel and local generalized channel respectively. 
		$n$ indicates the number of times the noise is applied to the quantum states.\\}
\end{table}

\subsubsection{Inherent Noise on Measurement}\label{inm}

%Quantum measurement is not merely a passive reading of information 
%but an active disturbance of the quantum system~\cite{ozawa2004uncertainty}.
%The interaction between the measurement device and the qubits causes the qubits' state to change,
%known as wave function collapse~\cite{sandhu2019enhancing}.
%After quantum measurement, 
%the quantum state indeed collapses to either 
%$|1 \rangle$  or  $|0 \rangle$,
%determined by the probability amplitudes of the superposition state before measurement. 
%This process is inherently random
%and this inherent randomness leads to the unpredictability of measurement results.
%In order to improve measurement accuracy,
%multiple measurements are often performed 
%to obtain a more precise estimate of the quantum state~\cite{wecker2015progress}.
Although repeated measurements of a qubit can indeed 
provide a more precise estimate of the quantum state, 
due to time and cost constraints,
the number of measurements is always limited.
This limitation introduces randomness, 
known as shot noise~\cite{li2023differential}. 
Central limit theorem can be applied to model shot noise as having a Gaussian or normally distribution~\cite{zhao2024bridging},
inspiring work on utilizing shot noise to satisfy QDP~\cite{li2023differential,du2022quantum}.

By reducing the number of measurements to increase shot noise, 
\citet{li2023differential} achieved the desired privacy protection for quantum system. 
They used projection operators to measure quantum states 
by projecting them onto a subspace. 
To be specific, 
they defined adjacent quantum states using quantum \emph{hockey-stick divergence} 
and introduce the impact of shot noise. 
Through both theoretical derivations (\Cref{the:3}) and numerical experiments, 
they demonstrated that increasing the number of measurements
increases the privacy budget of QDP, 
which means the level of privacy protection decrease.
Besides, 
\cite{du2022quantum} designed an efficient QDP Lasso estimator, 
which also increases shot noise (randomness) 
by reducing the number of quantum measurements. 
They believed that reducing the number of measurements to achieve a specific privacy budget is simpler
compared to introducing additional noise into the quantum circuit,
thereby resulting in faster runtime.

\begin{theorem} Measurement Mechanism~\cite{li2023differential} \label{the:3}.
	Assume the number of measurements (shots) is $n$
	and the set of projection operators is $\{M_i\}$ 
	with maximum rank $r$ of $\{M_m\}$. 
	For all quantum states $\rho$ and $\sigma$ with $F(\rho, \sigma) \leq d$, 
	measurement mechanism hold ($\epsilon$, $\delta$)-QDP with
	\begin{equation}\label{equ:20}
		\epsilon_{m} = \frac{dr}{(1 - \mu)\mu} \left[ \frac{9}{2}(1 - 2\mu) + \frac{3}{2}\sqrt{n} + \frac{dr(\mu + dr)n}{1 - \mu} \right]
	\end{equation}
	where $\mu = \min\{\mu_1, \mu_0\}$ with
	$\mu_0 = \text{Tr}(\rho M_m)$ and
	$\mu_1 = \text{Tr}(\sigma M_m)$.
\end{theorem}

%\subsection{Hybrid QDP Realization}

\subsection{External Noise to Realize QDP}\label{sec-external}

The inherent noise in quantum computing can provide partial differential privacy protection, 
which is a natural advantage over classical computing.
However, 
achieving strict privacy protection usually requires the addition of external noise~\cite{yang2023improved}.
On one hand, 
when privacy requirements are high, 
the inherent noise in quantum computing cannot be sufficiently increased 
within limited resources, 
and thus may not meet the privacy protection requirements.
%~\cite{resch2021benchmarking}, 
On the other hand, 
although some studies have quantified the noise 
in quantum channels and provided probabilistic models~\cite{angrisani2022differential,angrisani2023unifying}, 
these quantifications are mainly theoretical, 
and the actual noise in practical operations is often uncontrollable~\cite{soare2014experimental}.
But artificially added noise, 
such as Laplacian noise in traditional DP, 
can be controlled and quickly adjusted according to the required level of privacy protection. 
Therefore, 
external noise needs to be supplemented in QDP, 
and it is generally added on quantum state
or quantum measurement after passing through the quantum circuit,
which are discussed in \Cref{enqs} and \Cref{enqm} correspondingly. 

\subsubsection{External Noise on Quantum State}\label{enqs}

When utilizing quantum computers to process private data, 
it is necessary to encode the data first, 
converting classical data into quantum states to perform quantum operations. 
This step also called \emph{data-encoding feature map}~\cite{pande2024comprehensive}.
Once mapped into quantum states, 
artificial noise can be added to these states to meet QDP requirements.
Due to the post-processing characteristics of QDP, 
the quantum states continue to meet QDP standards 
even after passing through the quantum circuit.

\citet{huang2023certified} adopted random rotation on quantum states to add noise.
%proposed a method to satisfy QDP 
%and enhance the robustness of quantum classifiers 
%using quantum random rotation noise. 
They applied random rotation gates (such as Pauli-X gates) on each qubit, 
with the random rotation angles. 
% They regarded random rotations on qubits as random smoothing.
They incorporated this key idea into 
the design of quantum classifiers, 
which add rotation noise to the input quantum states,
and demonstrated that such quantum classifiers satisfy QDP.
%then the quantum classifier was executed 
%on each quantum state with added random rotation noise 
%to obtain classification results. 
%They derived the relationship between QDP 
%and the magnitude of the added rotation noise,
%which is shown in \cref{comparisonex}.
\citet{gong2024enhancing} first applied random unitary matrix encoding to the input data to mask its details. 
Based on \cite{huang2023certified}, they added rotation noise to the quantum states to achieve QDP. 
After that, during the error correction stage, 
they use \emph{Quantum Error Correction} (QEC) encoding~\cite{fukui2017analog} 
to amplify the QDP and provide an amplification bound. 
Specifically, they proposed a black-box QEC encoder, 
unknown to attackers, 
which corrects errors while enhancing privacy.
%In addition,
%\citet{angrisani2022differential} proved that quantum encodings inherently offer approximate DP 
%and claimed that combining quantum encoding with 
%the Laplace and Gaussian mechanisms can amplify DP. 
%
\Cref{comparisonex} shows the comparison of these two work.

%
%\begin{theorem}\label{the:4} 
%Let the algorithm $M$ correspond to the class classification circuit with random rotation noise and measurement operators. 
%For two neighboring quantum states $\sigma$ and $\rho$ 
%satisfying $\tau(\rho, \sigma) \leq d $ and \(0 \leq d \leq 1\), 
%the algorithm $M$ satisfies $\epsilon$-QDP, where:
%\begin{equation}\label{equ:21}
%\epsilon = \ln \left(1 + \frac{d}{tn} \right)
%\end{equation}
%where \(t\) is the noise magnitude parameter, and \(n\) is the number of qubits.
%\end{theorem}

%\begin{table}[h!]
%	\caption{The comparison of external noise on qubits}
%	\centering\footnotesize
%	\label{comparisonex}
%	\begin{tabularx}{\columnwidth}{p{0.3\columnwidth}cp{0.5\columnwidth}}
	%		\toprule
	%		\textbf{Noise type} & \textbf{Quantum differential Privacy} & \textbf{Notations} \\
	%		\midrule
	%		Random rotation noise& $\ln \left(1 + \frac{d}{tn} \right)$-QDP & 
	%		$t$ is the noise magnitude parameter and $n$ is the number of qubits.\\
	%		B & 789 & 012 \\
	%		C & 345 & 678 \\
	%		\bottomrule
	%	\end{tabularx}
%\end{table}

\begin{table}[t] 
	\centering\scriptsize
	\caption{The comparison of external noise on qubits}
	\label{comparisonex}
	\resizebox{0.47\textwidth}{!}{
		\begin{tabular}
			{m{70pt}<{\centering}m{65pt}<{\centering}m{55pt}<{\centering}}
			\toprule
			\textbf{Aspect} &\textbf{\citet{huang2023certified}}& \textbf{\citet{gong2024enhancing}} \\
			\midrule
			Mechanism & Random rotation in quantum classifier& QEC Encoding in quantum classifier  \\ \midrule 
			Stage in quantum computing & Initialization & Error correction\\ \midrule
			Privacy Protection & Initial Privacy Protection & Privacy Amplification\\\midrule
			Privacy budget in QDP & $\ln \left(1 + \frac{d}{t^n} \right)$ &$\epsilon\left( \frac{n_0(n_0 - 1) d^2}{\delta} \right)$\\ \midrule[0.1pt]
			Noise controllable parameter &  Noise magnitude parameter $t$& The distance $d$\\
			\bottomrule
	\end{tabular}}
	\parbox{\columnwidth}{\scriptsize
		1. $n$ is the number of qubits\\
		2. $\epsilon$ is QDP parameter before privacy amplification with QEC encoding 
		and $n_0$ is the number of physical qubits each logical qubit is encoded into in QEC\\}
\end{table} 

%\begin{table}[htbp]
%	%\renewcommand{\arraystretcxh}{1.0}
%	\caption{The comparison of external noise on qubits}
%	\centering\footnotesize
%	\label{comparisonex}
%	%\centering
%	\resizebox{\columnwidth}{!}{
	%		\begin{tabular}{c c}
		%			\toprule
		%			\textbf{Noise} & \textbf{Stage in quantum computing} & \textbf{QDP} & \textbf{Noise controllable parameter} \\
		%%			& \textbf{Other notations} \\ 
		%			\midrule
		%			Random rotation noise~\cite{huang2023certified} & Initialization &$\ln \left(1 + \frac{d}{tn} \right)$-QDP~\textsuperscript{1}
		%			&  Noise magnitude parameter $t$\\
		%			QEC Encoding~\cite{gong2024enhancing}& Error correction &  $(\epsilon\left( \frac{n_0(n_0 - 1) d^2}{\delta} \right), \delta)-QDP~\textsuperscript{2}$ & the distance $d$\\
		%%%			
		%%			\multirow{1}{2.5cm}{\centering Random rotation noise} & 
		%%			\multirow{1}{5cm}{\centering $\ln \left(1 + \frac{d}{tn} \right)$-QDP} & 
		%%			\multirow{1}{5cm}{\centering $t$ is the noise magnitude parameter and $n$ is the number of qubits}
		%%			\vspace{5mm}\\
		%
		%			\bottomrule
		%		\end{tabular}
	%	}
%	\parbox{\columnwidth}{\scriptsize
	%		1. $n$ is the number of qubits\\
	%		2. $\epsilon$ is QDP parameter before privacy amplification with QEC encoding and $n_0$ is the number of physical qubits each logical qubit is encoded into in QEC\\}
%\end{table}

\subsubsection{External Noise on Quantum Measurement}\label{enqm}

From \Cref{the:3}, 
we can see that the level of privacy protection is related to the number of measurements,
so if attackers can measure the qubits multiple times, 
the measurement results can approximate the true qubit state, 
leading to provide little privacy protection~\cite{kandala2019error}. 
Therefore, 
some researches has been made to artificially add noise
to the qubits during measurement to enhance privacy protection~\cite{ aaronson2019gentle,angrisani2022quantum}. 
These noise sources stem from traditional DP mechanisms, 
such as \emph{Gaussian mechanism} or \emph{Laplace mechanism}.
More precisely,
the external noise added during the measurement stage make the quantum algorithm satisfy DP rather than QDP 
(here we refer to the mechanism of applying traditional DP noise 
to quantum computing as satisfying ``fake QDP'').
This is because these mechanisms, 
when applied to the qubits post-measurement, 
function similarly to noise injection in classical scenarios. 
They do not alter the intrinsic properties of the quantum states themselves 
but rather affect the original data derived from these quantum states.

\citet{aaronson2019gentle} proposed a novel method named \emph{Quantum Private Multiplicative Weights} (QPMW)
for quantum measurement which satisfy LDP. 
In QPMW, 
they extended the \emph{Laplace mechanism} to quantum measurements 
by adding laplace noise in each round measurement result, 
ensuring that the probability distributions of measurement outcomes for different quantum states 
satisfy the definition of LDP. 
They also presented the concept of \emph{gentle measurement} of quantum states,
which means changing the trace distance of the quantum state by no more than 
$\alpha$ after the measurement.
They proved that the addition of laplace noise in measurements
enables $\alpha$-gentle measurement.
Moreover, 
they use the \emph{randomized response} mechanism to replace the \emph{Laplace mechanism} in QPMW to protect quantum privacy. 
%It has been shown that if each measured qubit is flipped with a probability of $\beta$, 
%then $\ln \left( \frac{1 - 2\beta}{1 + 2\beta} \right)$-DP can be achieved.
Then, 
based on~\cite{aaronson2019gentle},
\citet{angrisani2022quantum} extended the quantum application of 
the  \emph{randomized response} to \emph{K-Randomized Response} (KRR).
They utilized \emph{Positive Operator-Valued Measure} (POVM) measurements~\cite{oszmaniec2017simulating}, 
which allow for outcomes that are not limited to traditional projective measurements 
and can have multiple non-orthogonal results. 
According to each POVM mesurement result, 
they specified the magnitude of the Laplace noise or KRR noise that should be added satisfy LDP requirements.

In contrast,
\citet{angrisani2022differential} chose to add laplace noise or gaussian noise 
to the mean of the measurement results for each encoded qubit.
They demonstrated how to achieve DP 
by combining quantum encoding with laplace or gaussian noise. 
In addition,
they suggested that \emph{subsampling} the measurement results 
can increase the privacy effect.
Following their previous work of adding noise to the mean of measurements, 
\citet{angrisani2023unifying} determined the noise parameters of 
the \emph{Gaussian mechanism} and \emph{Laplace mechanism} for quantum.
The process of two methods that add noise on qubits measurement
is shown in \Cref{fig:noise},
where $\rho$  and $\sigma$ are two neighboring qubits after encoding.
The comparison of different mechanism on quantum measurement  is shown in \Cref{measurement}.

\begin{figure}[htbp]
	\centering
	\includegraphics[width = \linewidth]{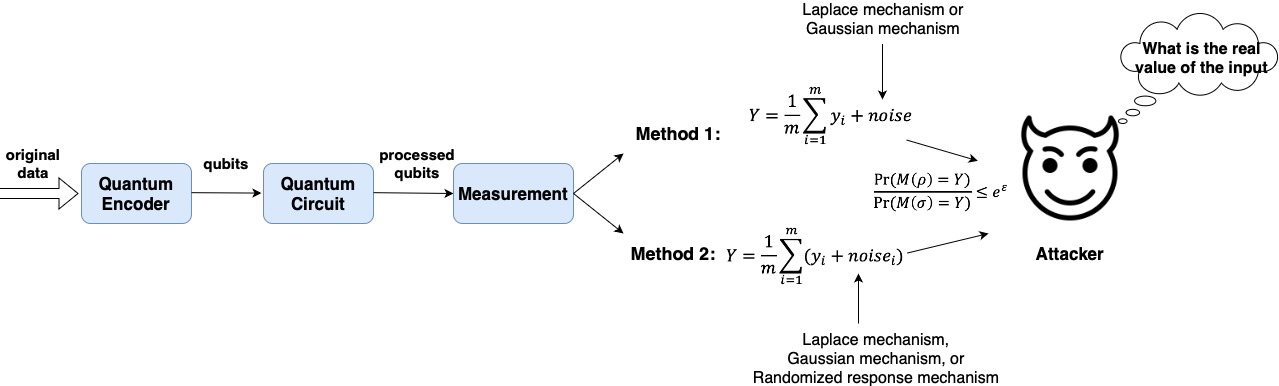}
	\caption{Two simple methods to add noise for quantum measurement with potential mechanisms where 
		1) Method \num{1} (DP):  add noise to the average value $\frac{1}{m} \sum_{i=1}^{m} y_i$ of $m$ rounds of measurement of a qubit 
		2) Method \num{2} (LDP): add noise to the each round of the measurement $y_i$ of a qubit }
	\label{fig:noise}
\end{figure}

\begin{table}[h]
	\centering\scriptsize
	\caption{The comparison of noise on quantum measurement}
	\label{measurement}
	\begin{tabular}{>{\centering\arraybackslash}m{3cm} >{\centering\arraybackslash}m{1.2cm} >{\centering\arraybackslash}m{1.7cm}>{\centering\arraybackslash}m{1cm}} 
		\toprule % Top horizontal line
		\textbf{Mechanism} & \textbf{Noise type} & \textbf{Privacy Framework}  & \textbf{References}\\ % Column names
		\midrule[0.1pt] % Middle horizontal line
		Measurement mechanism (shot noise) & Inherent noise & QDP &\cite{li2023differential} \\ \midrule
		Randomized response & External noise & LDP &\cite{aaronson2019gentle}\\ \midrule
		K-randomized response & External noise & LDP  &\cite{angrisani2022quantum}\\\midrule[0.1pt]
		Laplace mechanism& External noise &  DP or LDP &\cite{angrisani2022differential}, \cite{angrisani2022quantum} \\\midrule
		Gaussian mechanism & External noise & DP&\cite{angrisani2022differential} \\
		\bottomrule % Bottom horizontal line
	\end{tabular}
\end{table}

\subsection{Understanding QDP Mechanisms}\label{summarym}
{

We present a comprehensive gap analysis of the literature 
on leveraging noise to protect quantum data, 
highlighting the need for mechanisms operating directly on quantum states, 
such as quantum channels and rotations. 

Traditional DP mechanisms, 
like the Laplace mechanism and Randomized Response, 
perturb measurement results rather than quantum states, 
leaving a gap in robust privacy protection. 
QDP introduces noise directly into quantum channels and operations, 
effectively countering specific attacks like man-in-the-middle 
and side-channel attacks~\cite{fei2018quantum, ravi2024side}. 
This approach better safeguards quantum states compared to traditional DP mechanisms.

\color{black}
\vspace{-3pt}
\section{Proposal I: Hybrid QDP}\label{hybrid}

In quantum systems, noise arises from diverse sources, impacting qubit states and computations. While existing QDP mechanisms typically address a single noise source~\cite{zhou2017differential,bai2024quantum}, they often neglect the joint influence of multiple noise sources. Addressing this gap, the proposed Hybrid QDP mechanism leverages both quantum channel noise $\mathcal{N}_c$ and qubit measurement noise $\mathcal{N}_m$, optimizing privacy budgets under combined noise conditions.
\vspace{-5pt}
\subsection{Hybrid Privacy Budget}
The total privacy budget, $\epsilon_s$, is calculated as:
\begin{equation}\label{equ:21}
    \epsilon_s = 1 - \prod_{i=1}^n (1 - \epsilon_i),
\end{equation}
where $\epsilon_s$ ranges between \num{0} and \num{1} to ensure adequate privacy. Here, $n$ represents the number of noise sources. For example, with $n=2$, if $\epsilon_1 \to 0$, $\epsilon_s \approx \epsilon_2$, adapting seamlessly to scenarios dominated by a single noise source.

The proposed mechanism optimizes noise management by efficiently allocating privacy budgets and balancing noise levels. It provides significant privacy preservation benefits, as demonstrated through mathematical validation.
\vspace{-2pt}
\subsection{Monotonicity of $\epsilon_s$}
To prove monotonicity, consider the partial derivative:
\begin{equation}\label{equ:23}
    \frac{\partial \epsilon_s}{\partial \epsilon_k} = \prod_{i \neq k} (1 - \epsilon_i).
\end{equation}
Since $(1 - \epsilon_i) > 0$, it follows that:
\begin{equation}\label{equ:24}
    \frac{\partial \epsilon_s}{\partial \epsilon_k} > 0,
\end{equation}
indicating $\epsilon_s$ increases monotonically with respect to each $\epsilon_k$.
\vspace{-7pt}
\subsection*{Boundary Conditions}
The boundaries of $\epsilon_s$ are analyzed as follows:
\begin{equation}\label{equ:25}
    \prod_{i=1}^{n} (1 - \epsilon_i) = \exp\left( \sum_{i=1}^n \ln(1 - \epsilon_i) \right).
\end{equation}

\textbf{Case 1:} When $\epsilon_i \to 0$, using the Taylor expansion $\ln(1 - \epsilon_i) \approx -\epsilon_i$:
\begin{equation}\label{equ:28}
    \epsilon_s \approx 1 - \exp\left(-\sum_{i=1}^n \epsilon_i\right) \approx \sum_{i=1}^n \epsilon_i.
\end{equation}

\textbf{Case 2:} When $\epsilon_i \to 1$:
\begin{equation}\label{equ:29}
    \epsilon_s \to 1 - \exp\left(-\sum_{i=1}^n \epsilon_i\right) \approx 1.
\end{equation}

\textbf{Case 3:} When $\epsilon_k \to 1$ and $\epsilon_i \to 0$ for $i \neq k$:
\begin{equation}\label{equ:31}
    \epsilon_s \to 1.
\end{equation}

These results validate the monotonicity and boundary conditions of \Cref{equ:21}, demonstrating its robustness in capturing the joint privacy impact of multiple quantum noise sources. The Hybrid QDP mechanism provides a mathematically sound framework for balancing noise and privacy in quantum systems.

\vspace{-7pt}
\subsection{Experiments Evaluation}
A quantum circuit is designed to implement the Proposed Hybrid QDP \footnote{The Proposed Hybrid QDP refer to as resilient mechanism in this paper.}, where
the quantum state initially undergoes a transformation
through a gate affected by channel noise. 
Subsequently, 
the quantum state is measured, 
incorporating the shot noise.
In the experiments, 
fidelity is used to measure of the similarity between
the real output quantum state $\rho$ (with noise) 
and the ideal output quantum state $ \sigma$ (without noise),
indicating the impact of noise on qubits.

%In quantum systems, 
%utility is often represented by fidelity~\cite{bai2024quantum,hirche2023quantum}.
%As illustrated in \Cref{equ:22}, 
%fidelity ($F$) is a measure of the similarity between
%the real output quantum state $\rho$ (with noise) 
%and the ideal output quantum state $ \sigma$ (without noise),
%indicating the impact of noise on qubits.
%A total of two hundred runs of the quantum circuit were conducted 
%with the parameter setting $d=0.1$, 
%and the resulting plots are presented in \Cref{fig:comsum}, 
%\Cref{fig:sumchannel} and \Cref{fig:commea}.

\begin{equation}\label{equ:22}
	F(\rho, \sigma) =\left\|\sqrt{\rho} \sqrt{\sigma}\right\|_1^2
\end{equation}

%We designed a quantum circuit with two qubits, which includes a Hadamard gate and a Controlled-NOT gate,
%helping to create quantum entanglement and make the qubit enters a superposition state. 
%Quantum channel noise was applied to the Hadamard gate, 
%and the quantum circuit was executed a total of 200 times with the parameter set to $d=0.01$. 

We designed a quantum circuit with two qubits, incorporating a Hadamard gate and a Controlled-NOT gate to create quantum entanglement and superposition. 
Quantum channel noise was applied to the Hadamard gate, and the circuit was executed 200 times with $d=0.01$.

\subsubsection{Exploring the Optimal Privacy Budget Allocation}\label{opb}

We first conduct experiments to explore the optimal allocation of the privacy budget 
under different  channel noise combined with shot noise (from measurement). 
The goal is to establish a noise management strategy, 
ensuring robust privacy protection while achieving the best fidelity.
We use \Cref{equ:21}, 
the expressions for 
different QDP channel mechanisms (e.g., \Cref{equ:12}), 
and the measurement mechanism (\Cref{equ:20}) 
to explore the various quantum channels under measurement. 
\Cref{fig:comdep}, \Cref{fig:comgad}, and \Cref{fig:compf} show the results of the hybrid QDP 
using the depolarizing channel, the GAD channel, and the phase-flip channel, respectively.
\Cref{fig:compb} illustrates the fidelity achieved with different privacy budget allocation in the hybrid QDP.

\begin{figure}[t]
	\centering 
	\subfloat[The relationship between $\epsilon_s$, $\epsilon_{dep}$, and fidelity]{ 
		\label{fig:subfigd1} %% label for first subfigure
		\includegraphics[width=0.345\linewidth]                 {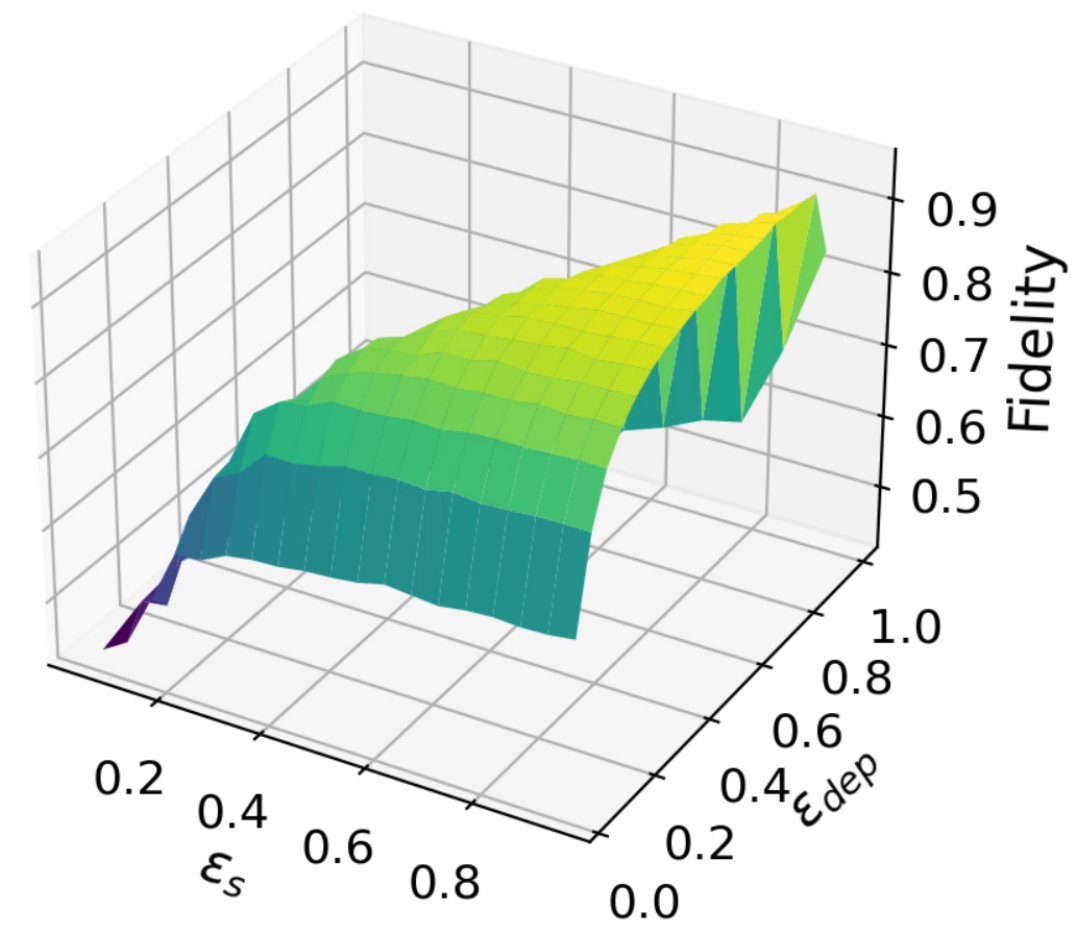}} 
	\hspace{5pt}
	\subfloat[\textcolor{black}{The relationship between $\epsilon_s$, $\epsilon_{m}$, and fidelity}]{ 
		\label{fig:subfigd2} %% label for second subfigure
		\includegraphics[width=0.345\linewidth]{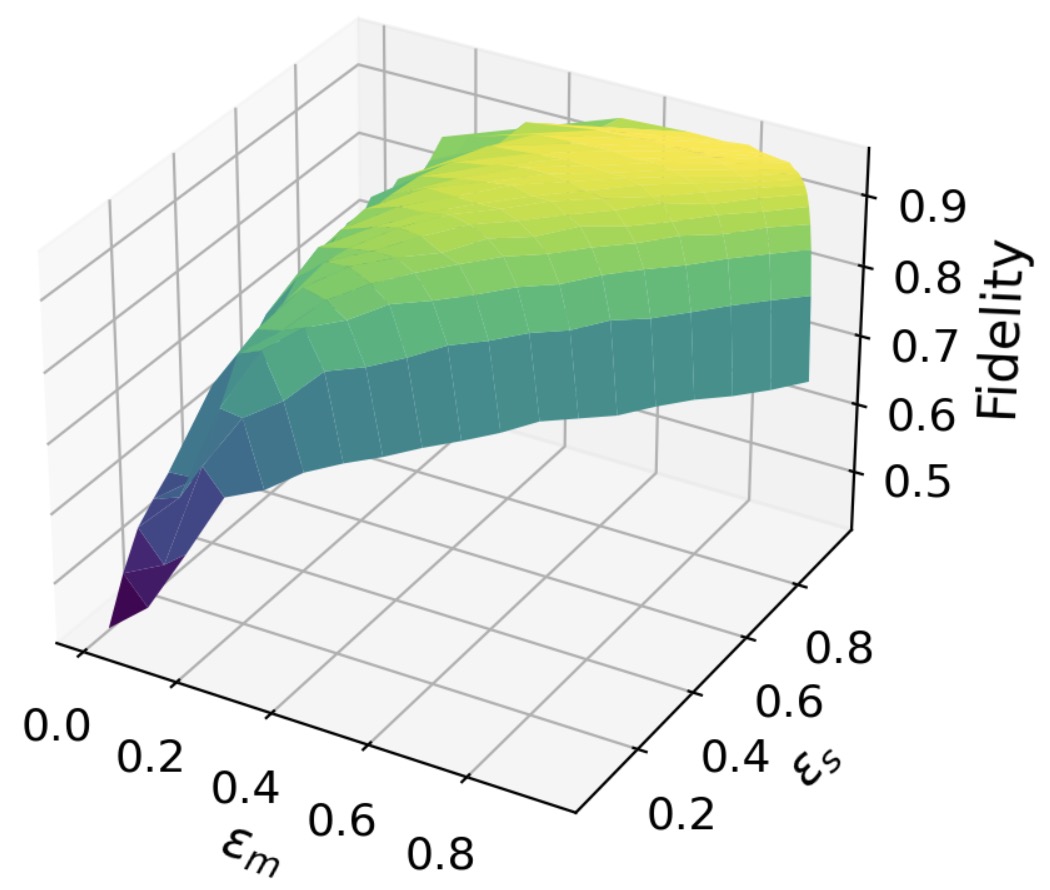}} 
	\vspace{5pt}
%	\subfloat[Phase-flip Channel with Measurement]{ 
%				\label{fig:subfigc3} %% label for first subfigure
%				\includegraphics[width=0.45\linewidth]{figure/pf3d.png}} 
	\caption{\textcolor{black}{The performance of depolarizing channel with measurement}} 
	\label{fig:comdep} %% label for entire figure 
\end{figure}

%\Cref{fig:subfigd1} depicts the relationship between 
%the fidelity of a qubit subjected depolarizing channel and measurement noise, 
%the total privacy budget ($\epsilon_s$), 
%and the privacy budget allocated to the depolarizing channel ($\epsilon_{dep}$). 
%\Cref{fig:subfigd2} shows the relationship between 
%the fidelity, 
%the total privacy budget ($\epsilon_s$), 
%and the privacy budget allocated to the measurement ($\epsilon_m$). 
From \Cref{fig:comdep}, 
it is evident that the fidelity increases with the increase of privacy budget 
(the privacy budget allocated to the measurement $\epsilon_m$, the privacy budget allocated to the depolarizing channel $\epsilon_{dep}$, or the total privacy budget $\epsilon_s$).
This is because the increase in the privacy budget results in a smaller amount of added noise to the qubit
and reduced privacy protection, 
thereby improving the fidelity.
Additionally, it can be observed that 
when the privacy budget ($\epsilon_m$, $\epsilon_{dep}$, or $\epsilon_s$) is less than \num{0.3}, 
the growth rate of fidelity is quite fast. 
After exceeding \num{0.3}, the growth rate slows down. 
This indicates that when the privacy budget is below \num{0.3}, 
the large noise amplitude has a significant impact on the quantum state.
%when $\epsilon_s$ is greater than 0.3, 
%the growth rate of fidelity slows down as $\epsilon_{dep}$ increases, 
%with the fidelity reaching a value of 0.75. 
%This indicates that when $\epsilon_s \leq 0.5$, 
%both $\epsilon_{dep}$  and $\epsilon_{m}$ have a significant impact on fidelity.
%However, when $\epsilon_s > 0.5$, 
%measurement noise had a  dominated impact on fidelity. At this point, to further improve fidelity, 
%the optimal approach is to increase the number of measurements.

\begin{figure}[htbp]
	\centering 
	\subfloat[\textcolor{black}{The relationship between $\epsilon_s$, $\epsilon_{gad}$, and fidelity}]{ 
		\label{fig:subfigg1} %% label for first subfigure
		\includegraphics[width=0.345\linewidth]                 {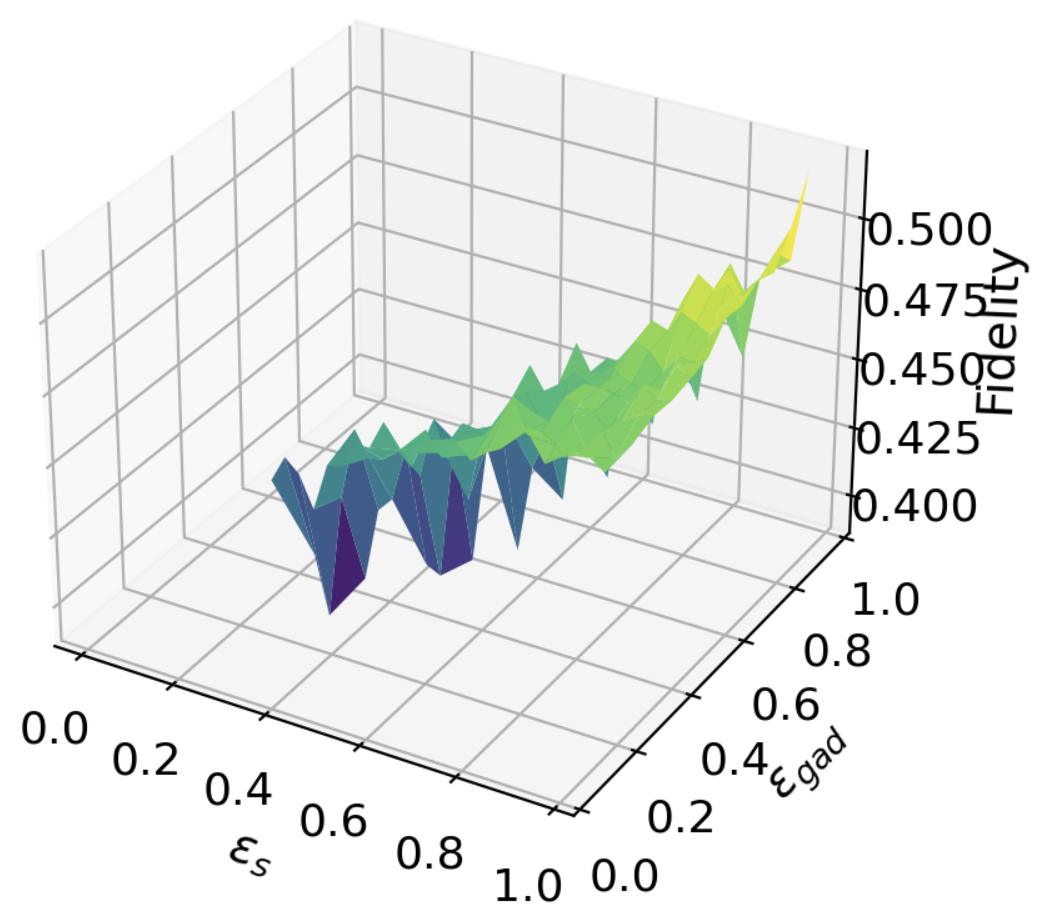}} 
	\hspace{5pt}
	\subfloat[\textcolor{black}{The relationship between $\epsilon_s$, $\epsilon_{m}$, and fidelity}]{ 
		\label{fig:subfigg2} %% label for second subfigure
		\includegraphics[width=0.345\linewidth]{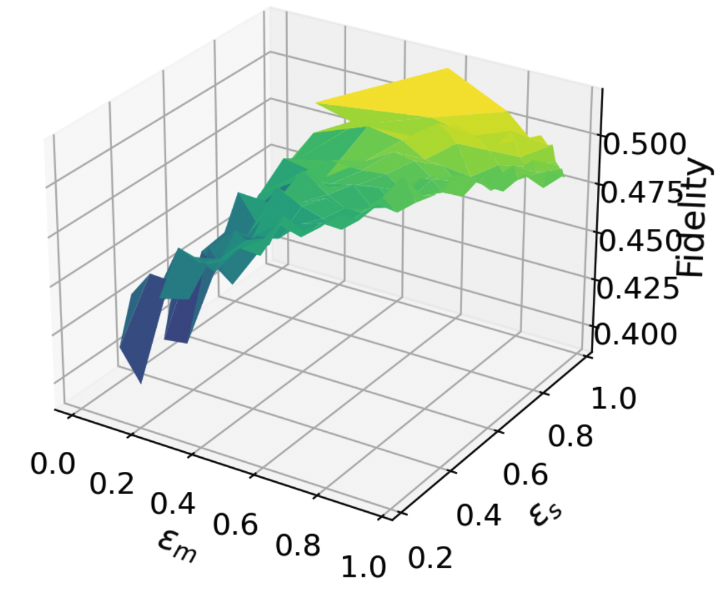}} 
	\vspace{5pt}
	%	\subfloat[Phase-flip Channel with Measurement]{ 
		%				\label{fig:subfigc3} %% label for first subfigure
		%				\includegraphics[width=0.45\linewidth]{figure/pf3d.png}} 
		\caption{\textcolor{black}{The performance of GAD channel with measurement}} 
		\label{fig:comgad} %% label for entire figure 
\end{figure}

%\Cref{fig:subfigg1} illustrates the relationship between 
%the fidelity of a qubit subjected GAD channel and measurement noise, 
%the total privacy budget ($\epsilon_s$), 
%and the privacy budget for GAD channel ($\epsilon_{gad}$). 
%\Cref{fig:subfigg2} shows the relationship between 
%the fidelity, 
%the total privacy budget ($\epsilon_s$), 
%and the privacy budget for measurement ($\epsilon_m$). 
Similarly, 
from \Cref{fig:comgad}, 
as $\epsilon_m$, $\epsilon_{gad}$ (the privacy budget for GAD channel), or $\epsilon_s$ increases, 
the fidelity also increases. 
%However, unlike the noise from the depolarizing channel, 
%when $\epsilon_s \leq 1$, 
%increasing $\epsilon_{gad}$ and $\epsilon_m$ both lead to a significant increase in fidelity. 
%Therefore, when $\epsilon_s < 1$, 
%if one wants to increase the fidelity, 
%any of the privacy budgets can be increased.
Compared to the noise from the depolarizing channel, 
the noise from the GAD channel appears to be more unstable, 
with larger fluctuations in \Cref{fig:comgad}. 
When $\epsilon_s$ reaches 1, 
higher values of $\epsilon_{gad}$ are more likely to achieve high fidelity (above 0.5).

\begin{figure}[t]
	\centering 
	\subfloat[\textcolor{black}{The relationship between $\epsilon_s$, $\epsilon_{pf}$, and fidelity}]{ 
		\label{fig:subfigpf1} %% label for first subfigure
		\includegraphics[width=0.345\linewidth]                 {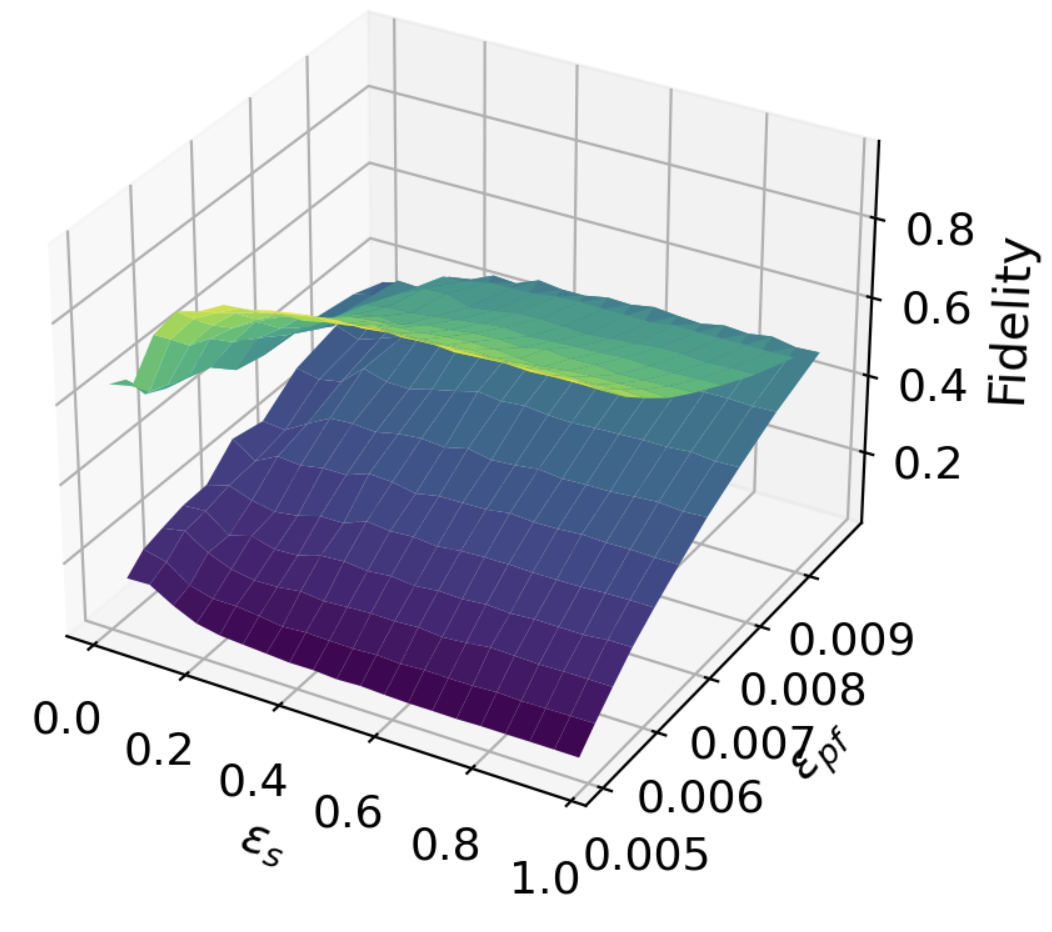}} 
	\hspace{5pt}
	\subfloat[\textcolor{black}{The relationship between $\epsilon_s$, $\epsilon_{m}$, and fidelity}]{ 
		\label{fig:subfigpf2} %% label for second subfigure
		\includegraphics[width=0.345\linewidth]{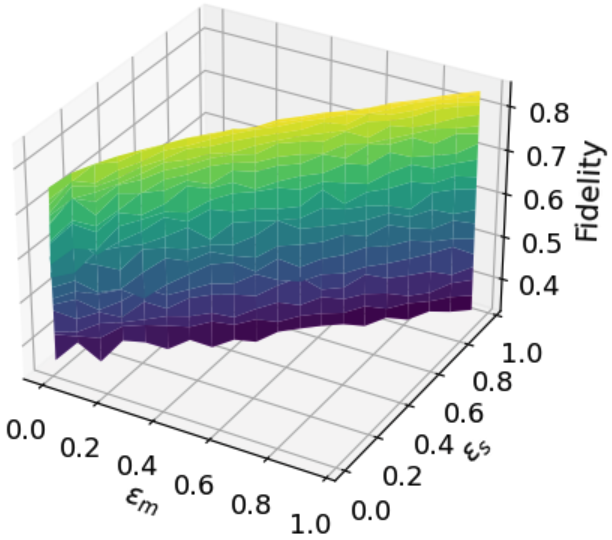}} 
	\vspace{5pt}
	%	\subfloat[Phase-flip Channel with Measurement]{ 
		%				\label{fig:subfigc3} %% label for first subfigure
		%				\includegraphics[width=0.45\linewidth]{figure/pf3d.png}} 
	\caption{\textcolor{black}{The performance of phase-flip channel with measurement}} 
	\label{fig:compf} %% label for entire figure 
\end{figure}

%\Cref{fig:subfigpf1} illustrates the relationship between 
%the fidelity of a qubit subjected phase-flip channel and measurement noise, 
%the total privacy budget ($\epsilon_s$), 
%and the privacy budget for phase-flip channel ($\epsilon_{pf}$). 
%\Cref{fig:subfigpf2} indicates the relationship between 
%the fidelity, 
%the total privacy budget ($\epsilon_s$), 
%and the privacy budget for measurement ($\epsilon_m$). 
One thing to note in \Cref{fig:compf} is that the maximum value of $\epsilon_{pf}$ (the privacy budget for phase-flip channel) is less than 0.01 
(which can be calculated based on the flip probability in the phase-flip channel).
It can be observed that for the same $\epsilon_{pf}$, there are two different fidelities. 
This is because $\epsilon_{pf}$ is symmetric with respect to the flip probability $p$, 
as shown in the expression for phase-flip channel in \Cref{comparisonchannel}.
$p$ is the probability that a qubit undergoes a flip, 
and $1-p$ is the probability that the qubit remains unchanged. 
Since $p$ and $1-p$ are symmetric, 
$\epsilon_{fp}$ is the same.
As $p$ increases, 
the probability of noise occurrence also increases, 
leading to a decrease in fidelity. 
Therefore, for the same $\epsilon_{pf}$, 
there are two different fidelities.
In the phase-flip channel, 
$p$ can be kept as small as possible (corresponding to a higher $\epsilon_{pf}$) 
to achieve a higher fidelity.

\begin{figure}[htbp]
	\centering 
	\subfloat[\textcolor{black}{Depolarizing Channel with Measurement}]{ 
		\label{fig:subfigpb1} %% label for first subfigure
		\includegraphics[width=0.28\linewidth] {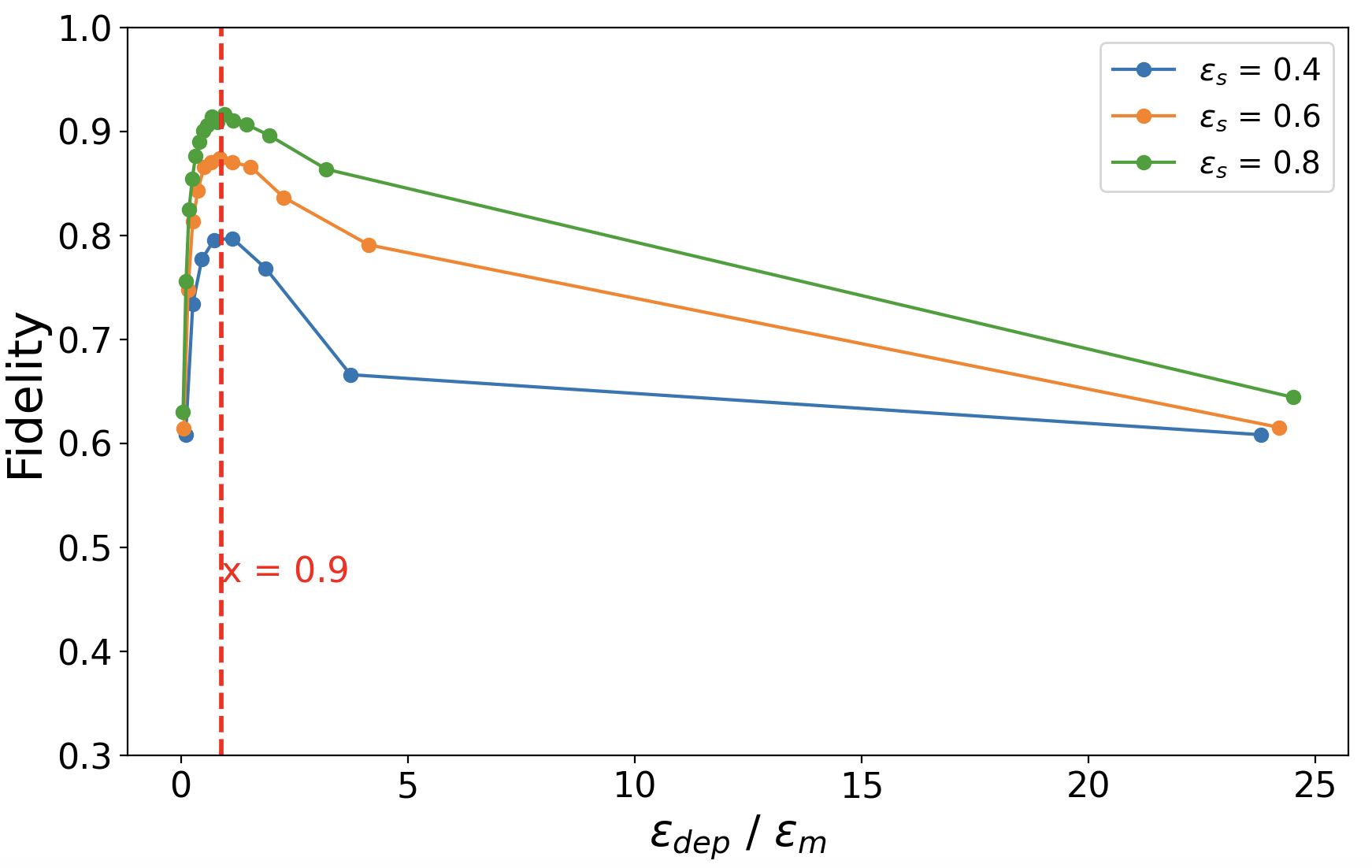}} 
	\hspace{5pt}
	\subfloat[\textcolor{black}{GAD Channel with Measurement}]{ 
		\label{fig:subfigpb2} %% label for second subfigure
		\includegraphics[width=0.28\linewidth]{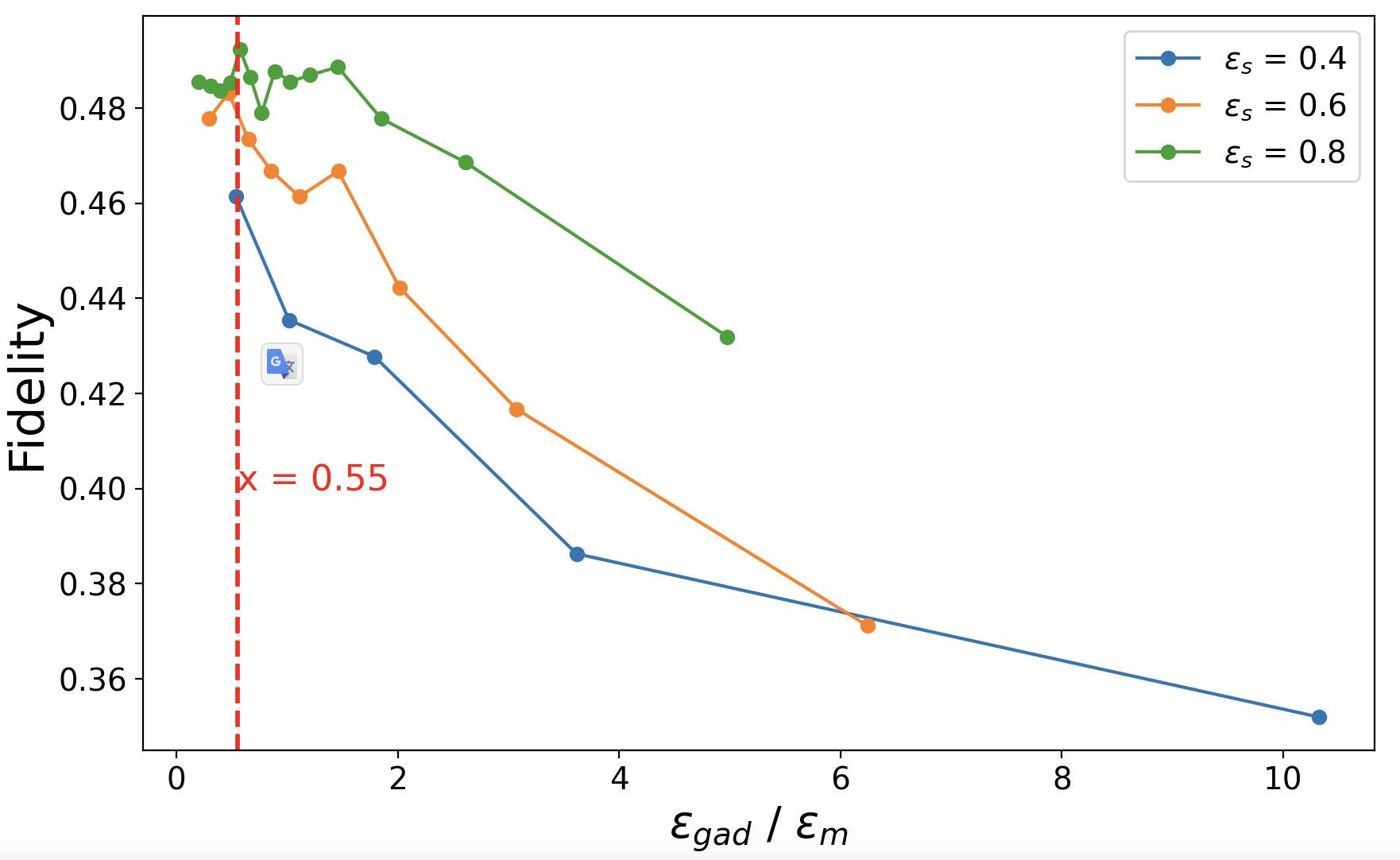}} 
	\vspace{5pt}
		\subfloat[\textcolor{black}{Phase-flip Channel with Measurement}]{ 
						\label{fig:subfigpb3} %% label for first subfigure
						\includegraphics[width=0.28\linewidth]{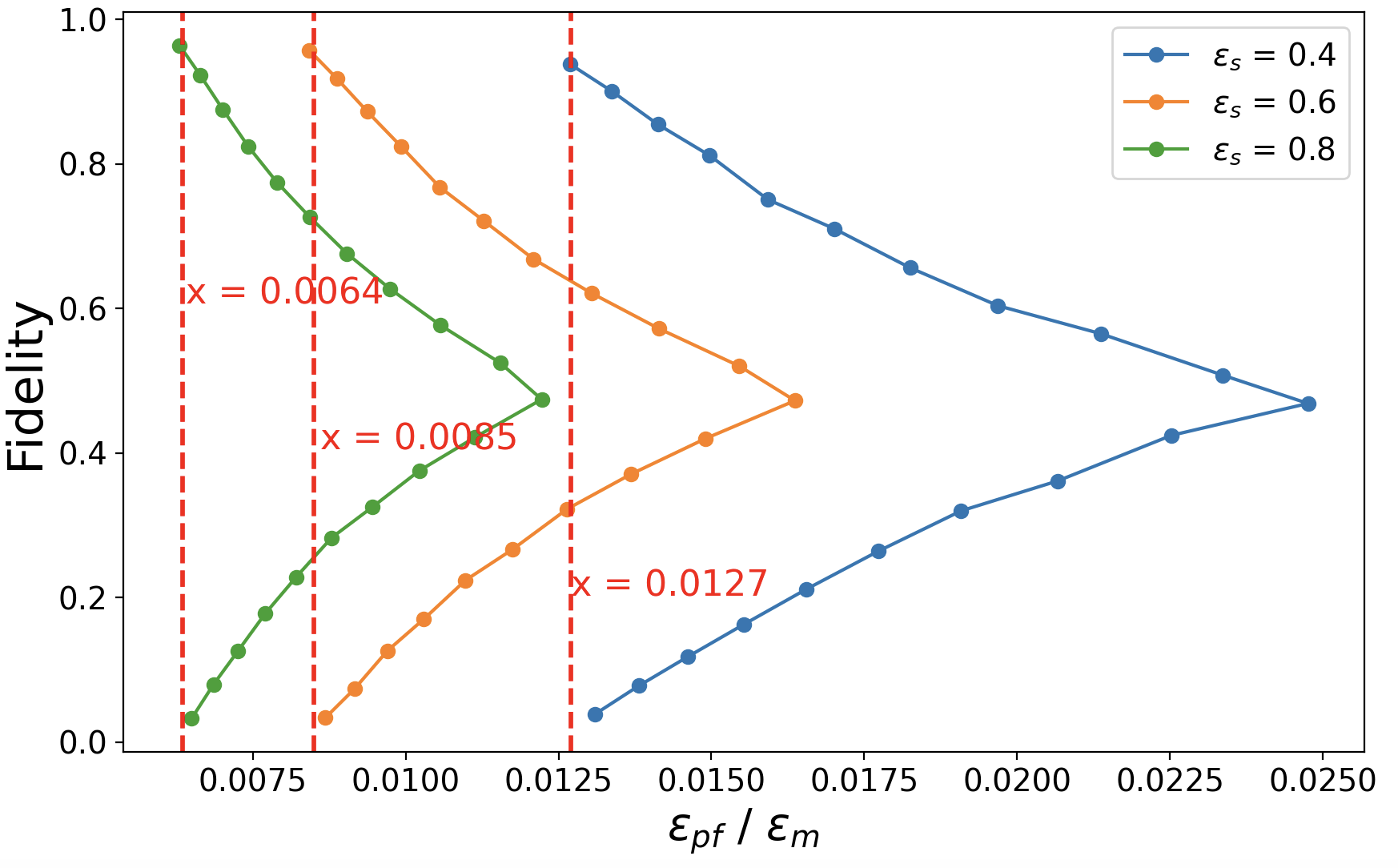}} 
	\caption{\textcolor{black}{The performance of phase-flip channel with measurement} }
	\label{fig:compb} %% label for entire figure 
\end{figure}

To explore the optimal allocation of the privacy budget between the channel and measurement noise, 
\Cref{fig:compb} shows the fidelity for different ratios 
of privacy budget for the channel and for measurement.
For the depolarizing channel and GAD channel, 
we can generally observe that for different values of $\epsilon_{sum}$, 
increasing the ratio of privacy budget allocated to the channel can  initially improve the fidelity, 
but after a point, it starts to decrease the fidelity. 
This is because allocating a smaller privacy budget to the channel or measurement 
causes larger disturbances to the qubit, 
resulting in lower fidelity. 
Therefore, it is important to choose an appropriate allocation ratio.
For the depolarizing channel with measurement, 
the ratio is approximately 0.9 to achieve the highest fidelity, 
while for the GAD channel with measurement, 
the ratio is approximately 0.55.
%This suggests that a larger portion of privacy budget should be allocated to the channel to add less noise, 
%while more noise should be introduced in the measurement (with fewer measurement) to achieve the optimal result. 
%This is because the quantum state first passes through the channel 
%and then the measurement. If the channel privacy budget is small, 
%the quantum state is already significantly affected, and even if the measure.
%In addition, as shown in \Cref{fig:subfigpb1} and \Cref{fig:subfigpb2}, 
%we can see that when the ratio is large, 
%a moderate increase in the ratio causes a decrease in fidelity. 
%This indicates that in such extreme cases, 
%excessive measurement noise is introduced, 
%and a small number of measurements (e.g., less than five times) leads to too much noise, 
%which affects the fidelity. 
%Therefore, for the depolarizing channel and measurement, 
%the optimal privacy budget ratio is around $\num{0.27}$, 
%while for the GAD channel and measurement, 
%the optimal privacy budget ratio is around $\num{2}$.
%One thing to note is that when $\epsilon_{sum}=1$, 
%the value of $\epsilon_{gad}/\epsilon_m$ cannot reach 2. 
%Therefore, the maximum attainable value is \num{1}.
For the phase-flip channel, 
due to the symmetry of its privacy budget, 
a small ratio can be considered. 
However, it is important to note that a small ratio will result in two fidelities. 
We need to ensure that both the ratio is small and $\epsilon_{pf}$ is also small 
in order to achieve a high fidelity.

\begin{figure}[htbp]
	\centering 
	\subfloat[\textcolor{black}{The relationship between $\epsilon_{m}$, $\epsilon_{gad}$, and fidelity}]{ 
		\label{fig:subfig31} %% label for first subfigure
		\includegraphics[width=0.345\linewidth] {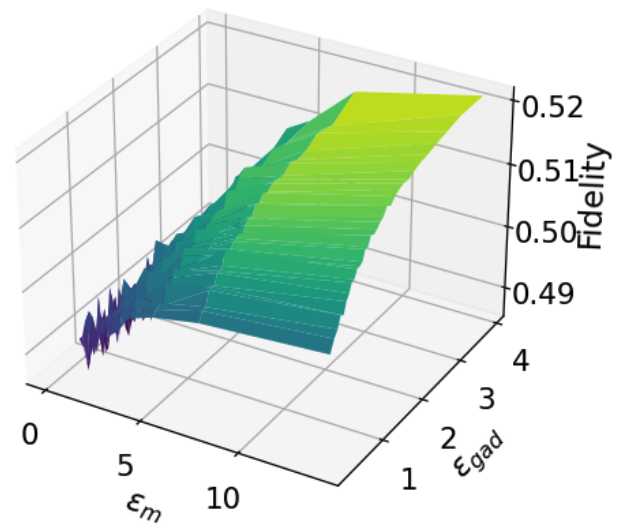}} 
	\hspace{5pt}
	\subfloat[\textcolor{black}{The relationship between $\epsilon_{gad}$, $\epsilon_{dep}$, and fidelity}]{ 
		\label{fig:subfig32} %% label for second subfigure
		\includegraphics[width=0.345\linewidth]{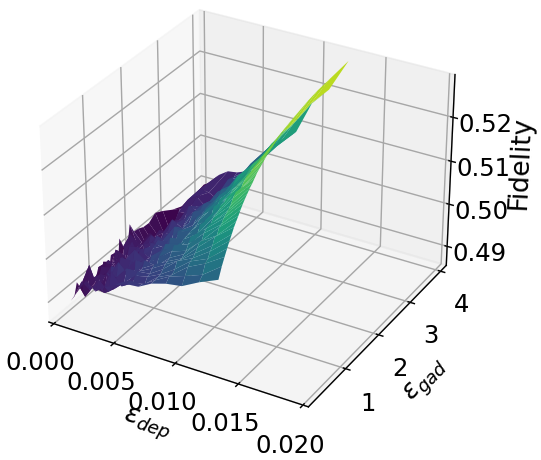}} 
	\caption{\textcolor{black}{The performance of depolarizing channel and GAD channel  with measurement}} 
	\label{fig:comp3} %% label for entire figure 
\end{figure}

Next, we investigate the optimal allocation of the privacy budget among three types of noise.
We assume that the quantum state is affected by depolarizing noise, generalized amplitude damping (GAD) noise, and shot noise (measuremnet).
We explore how to distribute the privacy budget, given that $\epsilon_{sum}=1$, 
to achieve the highest fidelity. 
The simulation results are shown in \Cref{fig:comp3}.
From \Cref{fig:subfig31} and  \Cref{fig:subfig32}, 
it is evident that the optimal privacy budget allocation for maximizing fidelity 
involves assigning the majority of noise to the depolarizing channel (small privacy budget), 
with smallest portions allocated to the measurement. 
This is because, under the same privacy budget, 
the fidelity of the depolarizing channel is the highest, 
as previously demonstrated by \cite{bai2024quantum}. 
Therefore, for the channel and measurement processes, 
to achieve high fidelity, 
the majority of the noise should be allocated to the channel, 
with the depolarizing channel receiving the most noise 
(the smallest privacy budget).

\subsubsection{Comparison with Single Noise Mechanism}\label{csn}

We compared the performance of the hybrid mechanism with single noise source mechanisms. 
It is important to note that the motivation for introducing the hybrid mechanism is to better simulate real-world conditions
(in a real quantum environment, having only one noise source is unrealistic), 
rather than necessarily achieving higher fidelity. 
In \Cref{fig:subfigs1}, 
we compare the hybrid mechanism 
(which includes noise from both the depolarizing channel and the measurement) 
with the depolarizing mechanism and the measurement mechanism. 
we can see that for small values of $\epsilon$, 
the depolarizing channel performs the best 
and the hybrid mechanism has the lowest fidelity. 
As $\epsilon$ increases, 
the gap between them gradually narrows, 
and by the time $\epsilon$ reaches 1, 
their fidelities essentially converge to the same value.
In \Cref{fig:subfigs2}, 
the hybrid mechanism includes noise from both the GAD channel and the measurement process, 
so we compare it with the GAD mechanism and the measurement mechanism. 
We observe that the fidelity of the hybrid mechanism is consistently lower 
than that of the single noise mechanisms. 
This may be due to the combined effect of GAD noise and shot noise, 
which leads to a greater disturbance on the quantum state.

\begin{figure}[htbp]
	\centering 
	\subfloat[\textcolor{black}{Hybrid mechanism with depolarizing channel}]{ 
		\label{fig:subfigs1} %% label for first subfigure
		\includegraphics[width=0.345\linewidth]                 {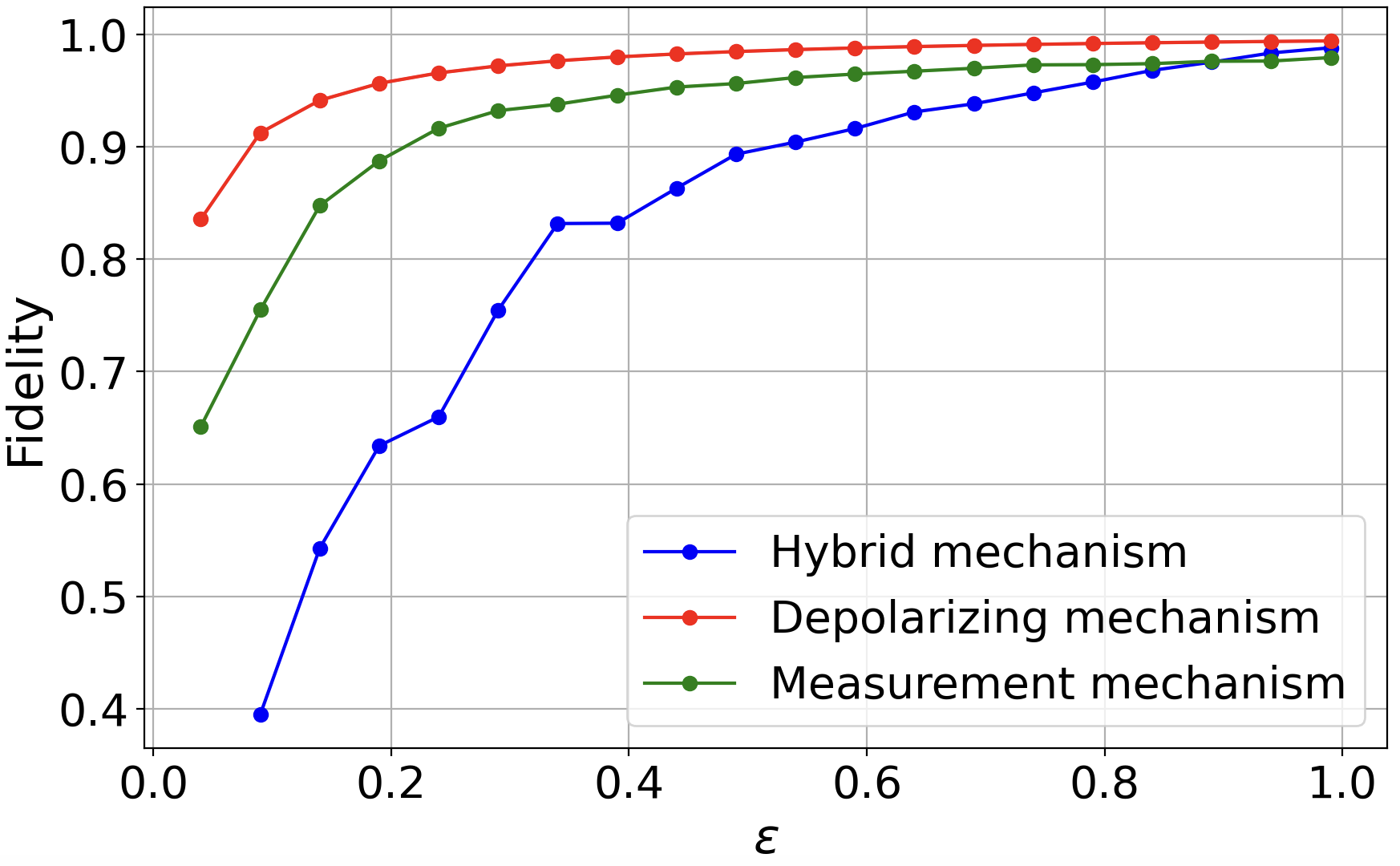}} 
	\hspace{5pt}
	\subfloat[\textcolor{black}{Hybrid mechanism with GAD channel}]{ 
		\label{fig:subfigs2} %% label for second subfigure
		\includegraphics[width=0.345\linewidth]{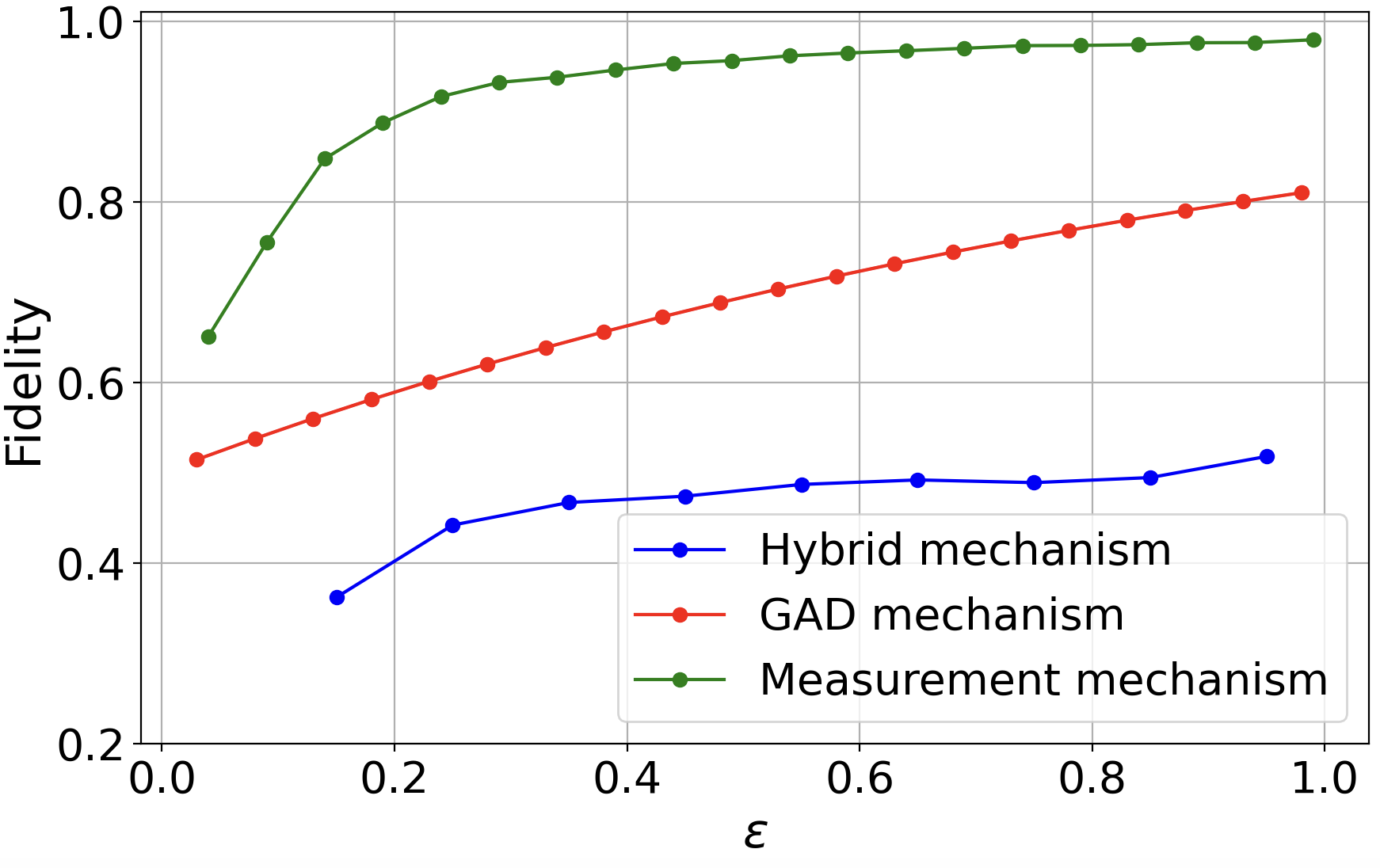}} 
%	\vspace{5pt}
%	\subfloat[Hybrid mechanism with phase-flip Channel with Measurement]{ 
%		\label{fig:subfigs3} %% label for first subfigure
%		\includegraphics[width=0.45\linewidth]{figure/hybridpf.jpg}} 
	\caption{\textcolor{black}{The comparison of hybrid mechanisms with others}} 
	\label{fig:comsi} %% label for entire figure 
\end{figure}

For the hybrid mechanism that includes phase-flip and measurement noise, 
we present its comparison results in \Cref{compf}
as the range of $\epsilon$ for the Phase-flip mechanism is between 0 and 0.01, 
while for the hybrid and measurement mechanisms, the range from 0 to 1. 
The Phase-flip mechanism achieves the highest fidelity (0.986) at $\epsilon=0.005$, 
significantly outperforming the other two mechanisms. 
Both the hybrid mechanism and the measurement mechanism reach their highest fidelity $\epsilon=1$, 
with the hybrid mechanism achieving a peak fidelity of 0.989.

\begin{table}[h]
	\centering\scriptsize
	\caption{\textcolor{black}{The comparison of hybrid mechanisms with Phase-flip mechanism and measurement mechanism}}
	\label{compf}
	\begin{tabular}{>{\centering\arraybackslash}m{2cm} >{\centering\arraybackslash}m{3cm} >{\centering\arraybackslash}m{3cm}} 
		\toprule % Top horizontal line
		\textbf{Mechanism} & $\epsilon=0.005$& $\epsilon=1 $\\ % Column names
		\midrule % Middle horizontal line
		Phase-flip mechanism& Fidelity= 0.986 & Nan  \\ \midrule[0.1pt]
		Measurement mechanism & Fidelity = 0.541 & Fidelity = 0.985 \\ \midrule[0.1pt]
		Hybrid mechanism & Fidelity = 0.632 & Fidelity = 0.989 \\
		\bottomrule % Bottom horizontal line
	\end{tabular}
\end{table}

\subsection{Findings and Insights}\label{summarym}

Recognizing that most existing mechanisms simulate only a single noise type, 
we developed a resilient mechanism encompassing both quantum channel noise and measurement noise. 
Our simulations included scenarios with two and three noise sources, 
and the results showed that more noise needs to be allocated to the channel rather than measurement 
(the optimal epsilon budget allocation ratio is less than 1).
Besides,
under the same privacy budget for channel noise, 
the depolarizing channel exhibited the highest fidelity. 
Thus, in practical applications, 
a larger privacy budget should be allocated to 
mitigating quantum channel noise to optimize quantum data utility, 
and when dealing with channel noise, 
the depolarizing noise type should be prioritized. 
While current QDP research is theoretical, 
further investigation and development are imperative. 
Effective noise management and control must be implemented in practical applications 
to realize QDP's full potential in real-world scenarios, 
ensuring robust privacy protections and secure quantum computing applications.

%\blindtext

%\lipsum[1]

%\begin{equation}
%    \epsilon=\epsilon_m + (1-\epsilon_m)\epsilon_d
%\end{equation}

\color{black}
\section{Proposal II: Lifted QDP Approach}\label{liqdp}

QDP primarily focuses on algorithm outputs from fixed quantum datasets, 
leveraging inherent noise to ensure privacy. 
However, it faces challenges in complex scenarios involving dynamic datasets, distributed quantum data, or quantum systems, due to limited adaptability and flexibility. Auditing QDP mechanisms is also resource-intensive, requiring repeated evaluations, which increases computational costs and sample complexity, particularly for dynamic quantum data.

To address these limitations, we propose \textit{Lifted Quantum Differential Privacy} (Lifted QDP) with relevant insights from~\cite{pillutla2024unleashing}. 
Lifted QDP introduces randomization to quantum datasets and rejection sets, 
enhancing QDP’s applicability and resistance to privacy breaches. 
This randomization allows the use of randomized canaries (here refer quantum canaries), 
rather than fixed deterministic examples (qubits). 
By leveraging statistical hypothesis testing to measure differences in outcomes, 
this approach significantly enhances the efficiency and scope of privacy auditing.
The definition is as follows:

\begin{definition} \label{def:liqdp} \textbf{Lifted Quantum Differential Privacy}.  
Let $P$ represent a joint probability distribution over ($\rho$, $\sigma$, $R$), where ($\rho$, $\sigma$) is a pair of random neighboring quantum states and $R$ is a rejection set. A quantum randomized mechanism $A$ satisfies $(\epsilon, \delta)$-Lifted QDP if, for any quantum measurement $M$, the following condition holds:
\begin{equation}\label{equ:10}
    \mathbb{P}_{A,P}[M(A(\rho)) \in R] \leq e^{\epsilon} \cdot \mathbb{P}_{A,P}[M(A(\sigma)) \in R] + \delta.
\end{equation}
\end{definition}

Given the definitions above, standard QDP is a special case of Lifted QDP. When the lifted joint distribution $P$ is independent of the algorithm $A$, Lifted QDP and QDP become equivalent. In Lifted QDP, the joint distribution $P$ represents the relationship between neighboring quantum states ($\rho$, $\sigma$) and the output of the algorithm $A$. If $P$ is independent of $A$, it implies that $P$ does not influence how $A$ processes quantum data. This independence ensures that the output of $A$ under Lifted QDP matches that under standard QDP, demonstrating that the lifting operation does not alter the algorithm's randomness or data protection capabilities.

\begin{table}[h]
	\centering\scriptsize
	\caption{\textcolor{black}{Comparison between QDP and Lifted QDP}}
	\label{comscaled}
	\color{black} 
	\begin{tabular}{>{\centering\arraybackslash}m{2cm} >{\centering\arraybackslash}m{3cm} >{\centering\arraybackslash}m{3cm}} 
		\toprule % Top horizontal line
		\textbf{Aspects} & \textbf{QDP} & \textbf{Lifted QDP}\\ % Column names
		\midrule % Middle horizontal line
		Scope of Randomization& Add noise on quantum states & Extend the scope of randomization to include quantum datasets, quantum states, and rejection sets  \\ \midrule[0.1pt]
		Protection Against Single Point Leakage & Higher risk of privacy loss if a quantum state is leaked & Provide stronger protection against single-point leakage \\ \midrule[0.1pt]
		Data Adaptability & Mainly applicable to static or fixed quantum data &  Better suited for handling dynamic quantum data, adaptable to changing datasets \\ \midrule[0.1pt]
		Flexibility& Less flexible, with fixed quantum datasets limiting adaptability & Greater flexibility, allowing dynamic adjustment of datasets and rejection sets\\
		\bottomrule % Bottom horizontal line
	\end{tabular}
\end{table}

The noise and mechanisms described in \Cref{sec-noise} and \Cref{sec-methods} for QDP are also applicable to Lifted QDP. Lifted QDP introduces randomization through quantum states or operations, extending noise application beyond individual operations to entire quantum datasets. This flexibility enables noise optimization across different states, allowing model reuse in multiple quantum tests, reducing sample complexity, and supporting repeated statistical testing. Additionally, Lifted QDP dynamically adjusts noise addition mechanisms for varying datasets, making it suitable for complex environments. The reuse of quantum states and operations across tests further enhances efficiency and reduces resource demands.

QDP auditing involves injecting a single canary (qubit) 
and observing the model's sensitivity to changes in the input data
to determine whether privacy is violated. 
However, this approach suffers from low information efficiency, 
as a single canary provides limited insight per experiment, 
requiring a larger number of samples (e.g., 1,000 experiments) 
to achieve stable statistical results. 
In contrast, Lifted QDP employs multiple randomized canaries (e.g., 32) during training, 
significantly reducing the number of experiments needed to achieve the same level of auditing accuracy. 
By requiring fewer samples (fewer auditing experiments), 
Lifted QDP enables a more efficient and effective auditing process.

\color{black}

\section{Challenges and Future Research Directions}\label{sec-future}

This section identifies key challenges and unanswered research directions in QDP, emphasizing practical limitations and proposing strategies to advance QDP from theory to real-world applications. Developing innovative solutions will enhance QDP's robustness, ensuring it becomes a reliable tool for safeguarding privacy in quantum computing.

\textit{DP Variants for Quantum Computing}\label{sec-variants}. QDP adapts traditional DP principles for quantum information processing. Unique quantum properties, such as superposition and entanglement, necessitate tailored variants. For instance, Quantum Local DP (QLDP)~\cite{angrisani2022quantum} addresses multiparty computations, while Quantum Pufferfish Privacy (QPP)~\cite{nuradha2024quantum} offers greater adaptability. Further research is needed to adapt these and other variants, such as Quantum Dynamic DP~\cite{zhang2016dynamic}, to quantum data's complex properties.

\textit{Privacy-Utility Trade-Off in QDP Mechanisms}\label{sec-resilient}. Current mechanisms often fail to account for multiple noise sources in quantum systems~\cite{angrisani2022differential,li2023differential}. Addressing noise interactions and optimizing privacy budgets are critical. Incorporating quantum error correction (QEC) and classical methods, such as machine learning for noise mitigation, could improve data utility and reliability in quantum computing.

\textit{Detecting Breaches in Noisy Algorithms}\label{sec-vulnerability}. Quantum system noise is often viewed as a natural privacy barrier. However, frameworks like~\cite{guan2023detecting} highlight the need to assess noisy algorithms' vulnerability to breaches. Future work should enhance the robustness of QDP algorithms and develop real-time monitoring systems to detect and prevent privacy violations, including applications of Lifted QDP introduced in \Cref{liqdp}.

\textit{Quantum Privacy Based on Hypothesis Testing}\label{sec-ht}. Quantum hypothesis testing offers privacy guarantees with explicitly defined error rates~\cite{farokhi2023quantum}. It complements QDP by providing operational insights and flexibility in privacy mechanisms. Future research should unify QDP and hypothesis-testing approaches, analyze robustness in noisy environments, and explore error probabilities as operational metrics.

\textit{Practical Implementation Strategies for QDP}\label{sec-practical}. Bridging the gap between theory and practice requires robust quantum error correction, noise management techniques, and customization of QDP protocols for specific hardware. Allocating resources to real-world deployment and iterative evaluation will ensure QDP solutions are effective and practical.

%%1. Combining Mechanisms to achieve QDP
%
%\cite{li2023differential}: the combination of depolarizing channel and measurement 
%
%\cite{angrisani2022differential}: the combination of depolarizing  channel and PAD, encoding and sampling
%
%2. Differential Privacy Variants in Quantum Computing
%
%3. QDP with quantum adversarial robustness
%
%4.summarize: Quantum Pufferfish Privacy: A Flexible Privacy Framework for Quantum Systems
%based on QDP

%1. Combining Mechanisms to achieve QDP
%
%\cite{li2023differential}: the combination of depolarizing channel and measurement 
%
%\cite{angrisani2022differential}: the combination of depolarizing  channel and PAD, encoding and sampling
%
%2. Differential Privacy Variants in Quantum Computing
%%Auditing Differential Privacy in High Dimensions with the Kernel Quantum Rényi Divergence
%
%%MATHEMATICAL COMPARISON OF CLASSICAL AND QUANTUM
%% MECHANISMS IN OPTIMIZATION UNDER LOCAL
%% DIFFERENTIAL PRIVACY ---ldp
%
%
%3. QDP with quantum adversarial robustness
%
%4.summarize: Quantum Pufferfish Privacy: A Flexible Privacy Framework for Quantum Systems
%
%5.real dp in quantum?
%%Detecting Violations of Differential Privacy for Quantum Algorithms   检测是否满足dp

%Barycentric and Pairwise Rényi Quantum Leakage信息泄漏，可以通过加dp解决 例如去极化通道

%\blindtext

%\lipsum[1]

\vspace{-0.7em}
\section{Conclusion} \label{sec-conclusions}
We explore the Quantum Differential Privacy (QDP), focusing on quantum noise and its potential in ensuring privacy. This research profiles noise types suitable for QDP, introduces a hierarchical mechanism to handle channel and measurement noise, and provides optimal privacy budget configurations to balance utility and privacy. A new concept of Lifted QDP is proposed, enhancing auditing efficiency and reducing sample complexity. Although the theory is robust, the practical implementation in quantum systems remains a challenge for ongoing research.

\bibliography{yourbib}
\bibliographystyle{IEEEtranN}

\end{document}